# Uniaxial compression of crystalline HCP titanium: an atomistic modelling study of size effects

Fatemeh Safari [a,*], Konstantinos Konstantinou [a,*]

[a] Department of Mechanical and Materials Engineering, University of Turku, Vesilinnantie 5, 20500 Turku, Finland

[*]Corresponding authors:

konstantinos.konstantinou@utu.fi

fatemeh.f.safari@utu.fi

## Abstract

Understanding the deformation behaviour of titanium is important not only for technological advances associated with industrially-relevant applications, but also essential to achieve a fundamental understanding of the mechanical properties of the relevant alloys. In this computational study, molecular-dynamics simulations are employed to ascertain the impact of model size on the mechanical response of alpha-titanium under compression. The deformation behaviour of the crystalline models is investigated as a function of different system sizes (varied by four orders of magnitude, up to 32 million atoms), and strain rates (down to $10^8$ $s^{-1}$). The results show that the elastic properties remain independent of both system size and strain rate, whereas marked size effects emerge during plastic deformation. Increasing the system size of the titanium model reduces stress fluctuations, results in more homogeneous structural evolution, and stabilizes dislocation activity. Decreasing the applied strain rate requires correspondingly a larger system size to achieve equilibration and to ensure a stable behaviour for the simulated structure. The modelling results demonstrate that system size and strain rate are strongly coupled, and their combined effect governs the simulated deformation behaviour of the compressed crystalline material.

# 1. Introduction

Titanium and its alloys are well known materials for their superior properties, and they have been widely employed across a versatile range of industries. For example, in the biomedical field, they are the materials of choice for implants and surgical instruments due to their high biocompatibility and excellent corrosion resistance [1–3]. In the aerospace sector, titanium alloys are often selected because they combine high strength with low weight, and maintain their performance under elevated temperatures, which make them suitable for airframe components and engine parts [4]. Furthermore, their high strength-to-weight ratio and durability are properties account for the utilization of these alloys in automotive applications [5].

For any given material, it is essential to study its mechanical behaviour in order to select and design suitable compounds tailored to specific applications. For instance, pure Ti can appear in different crystal structures; at room temperature it is in the form of hexagonal close-packed (HCP) structure (α-Ti), while at temperatures above 882 °C it transforms into body-centered cubic (BCC) phase (β-Ti) [6]. Plastic deformation plays an important role in the processing of Ti alloys, since it enables significant enhancement of their mechanical performance [7]. In Ti and other HCP metals, plastic deformation occurs mainly by dislocation of slip, twinning, and phase transformations, mechanisms that accommodate localized internal stresses [8–10]. Therefore, a clear understanding of the underlying deformation mechanisms is crucial for the development and improved functionality of these alloys.

The deformation behaviour of Ti has been the subject of numerous previous studies in order to clarify how slip and twinning contribute to plasticity under different loading conditions. Morrow et al. [11] investigated experimentally the dynamic loading of single crystal α-Ti, and observed that plastic deformation occurs from a combination of slip and twinning, with twinning being the dominant mechanism. In principle, the activation of specific twin variants inside the crystal structure is highly influenced by the corresponding crystallographic orientation. Bishoyi et al. [12], demonstrated that in the compression of commercially pure titanium, twinning activity depends on both temperature and crystallographic orientation. In particular, their experimental measurements showed that {10-12} twins were dominant at low and intermediate temperatures, while their contribution was reduced as the deformation progressed.

Alongside experimental studies revealing useful macroscopic behaviours, molecular-dynamics (MD) simulations have been previously employed to study the dynamical evolution of the deformation processes in crystalline Ti, and describe the relevant mechanisms at the atomic level [13,14]. MD simulations have been extensively applied to explore the mechanical behaviour of Ti across different microstructures, such as single-crystal [12,13,15], bicrystals [16], polycrystals [17–20], and nanowire structures [21], as well as under different deformation modes, including tensile [10,19,22], compressive [13,16,23], and shear loading [10,24]. These computer simulation studies have consistently supported the experimentally reported findings, demonstrating that the crystallographic orientation [25–27] and temperature [28,29] strongly influence the activation of deformation mechanisms. It was shown that in α-Ti, this arises from the limited set of slip systems, including basal <a>, prismatic <a>, and pyramidal <c+a> dislocations [30]. When these relevant slip systems cannot fully accommodate the applied strain, then deformation twinning emerges as a complementary mechanism to complete the deformation of the simulated system [30].

In atomistic simulation studies, in which critical phenomena are investigated, it is imperative to examine systematically the influence of finite-size effects on the calculated results [31]. The solidification process in Al, Fe, and Mg was explored with MD simulations, using model sizes ranging from ~2,000 up to ~8 million atoms, and it was shown that quantitative characteristics, such as nucleation time, nucleation rate, and free-energy changes, only became size-independent once the system reached approximately 2 million atoms [32]. This highlights that the simulation system size is not only a computational choice, but also a physical requirement for capturing meaningful aspects of the nucleation process and the defect-evolution behaviour in the modelled crystals. However, it should be noted that increasing the simulation system size rapidly increases the computational cost, and consequently, the detailed investigation of the dependence of the simulation results on size effects in studies of crystalline materials corresponds to an arduous task.

In principle, to describe realistically the plastic deformation process with MD simulations, the modelled system must be sufficiently large to suppress finite-size effects [33], and to allow the generation of key deformation defects, including dislocations and twins. The impact of finite-size effects on the deformation behaviour of modelled structures has been explored in MD studies of various material systems, such as nanopillars [23], nanowires [34], and polycrystals (grain size effects) [20]. Nevertheless, systematic investigations that isolate system size alone, under identical loading and microstructural conditions, remain limited.

Modelling the mechanical response of crystalline solids upon loading conditions provides a strong stimulus for performing computer simulations. In this study, we aim to achieve an understanding, at the atomistic level, of the influence of the simulation system size on the predicted mechanical behaviour of α-Ti under c-axis compression. We employ MD simulations for a wide range of modelled systems, created with an identical crystallographic orientation, exploring cell sizes up to 32 million atoms, and with utilizing the same interatomic potential and loading conditions for all the simulated structures.  In this way, we provide a coherent picture of the interplay between finite-size effects and the yield point, defect nucleation, and dislocation activities in the modelled structure of compressed α-Ti. In addition, four different strain rates, down to $10^8$ ($s^{-1}$), are employed for the investigation of the relevant processes in selected model system sizes to further examine the robustness of size-dependent trends in the compressed structures. We demonstrate how strongly the system size influences the reliability of MD simulations, and that large-scale simulations are necessary to capture realistic plastic deformation mechanisms during the c-axis compression of α-Ti.

# 2. Simulation methods

Initial crystalline α-Ti structures were generated using the Atomsk software [35]. The experimental lattice parameters at room temperature, a = 2.95 Å and c = 4.68 Å, [36] were chosen to build all the initial Ti models, with the x-axis, y-axis and z-axis aligned to the <10-10>, <1-210>, and <0001> directions, respectively [39]. An orthogonal hexagonal close-packed (hcp) supercell (with 4 atoms) was first constructed, and then different simulation boxes were created by replicating this supercell $N_x \times N_y \times N_z$ times (e.g., 10 × 10 × 10 for the smallest model studied here). In total, and based on this setup, seven model structures were generated, varying in size by four orders of magnitude with respect to the total number of atoms in each periodic simulation box, from 4,000 to 32 million atoms. The details regarding the name, the total number of atoms, the replication, and the extent of the simulation box in each dimension for all the Ti models studied here are summarized in Table 1. Moreover, in Fig. 1, three selected model structures from the different system sizes are displayed, highlighting the changes in the dimensions of the simulation box as the number of atoms increases.

Table 1. Details about the seven different Ti crystalline modelled systems in terms of the total number of atoms, the replication of the supercell, and the dimensions of the simulation box.

| **Model** | **Atoms** | **Replication (N)** | **$L_x$ (nm)** | **$L_y$ (nm)** | **$L_z$ (nm)** |
|---|---|---|---|---|---|
| 4k | 4,000 | 10 | 5.11 | 2.95 | 4.68 |
| 32k | 32,000 | 20 | 10.22 | 5.90 | 9.36 |
| 108k | 108,000 | 30 | 15.33 | 8.85 | 14.04 |
| 1M | 864,000 | 60 | 30.66 | 17.70 | 28.08 |
| 4M | 4,000,000 | 100 | 51.10 | 29.50 | 46.80 |
| 10M | 10,976,000 | 140 | 71.53 | 41.30 | 65.52 |
| 32M | 32,000,000 | 200 | 102.19 | 59.00 | 93.60 |

The compressive behaviour of α-Ti along c-axis (as depicted by the black arrow in Fig. 1) was studied by carrying out classical MD simulations, using the open-source LAMMPS (Large-scale Atomic/Molecular Massively Parallel Simulator) package [37]. An essential aspect associated with the simulations performed here corresponds to capturing plastic deformation in hcp Ti. In MD simulations, the accuracy of the modelled processes is highly dependent on the choice of the interatomic potential. Although several potentials have been widely used for Ti [38–43], a comparative study by Rawat et al. [44] demonstrated that the second-nearest neighbour modified embedded-atom method (2NN-MEAM) potential developed by Kim et al. [38] provides the most satisfactory performance under c-axis compression of Ti. Hence, this interatomic potential was chosen to model the interactions between the Ti atoms, and to study the relevant deformation processes in this work.

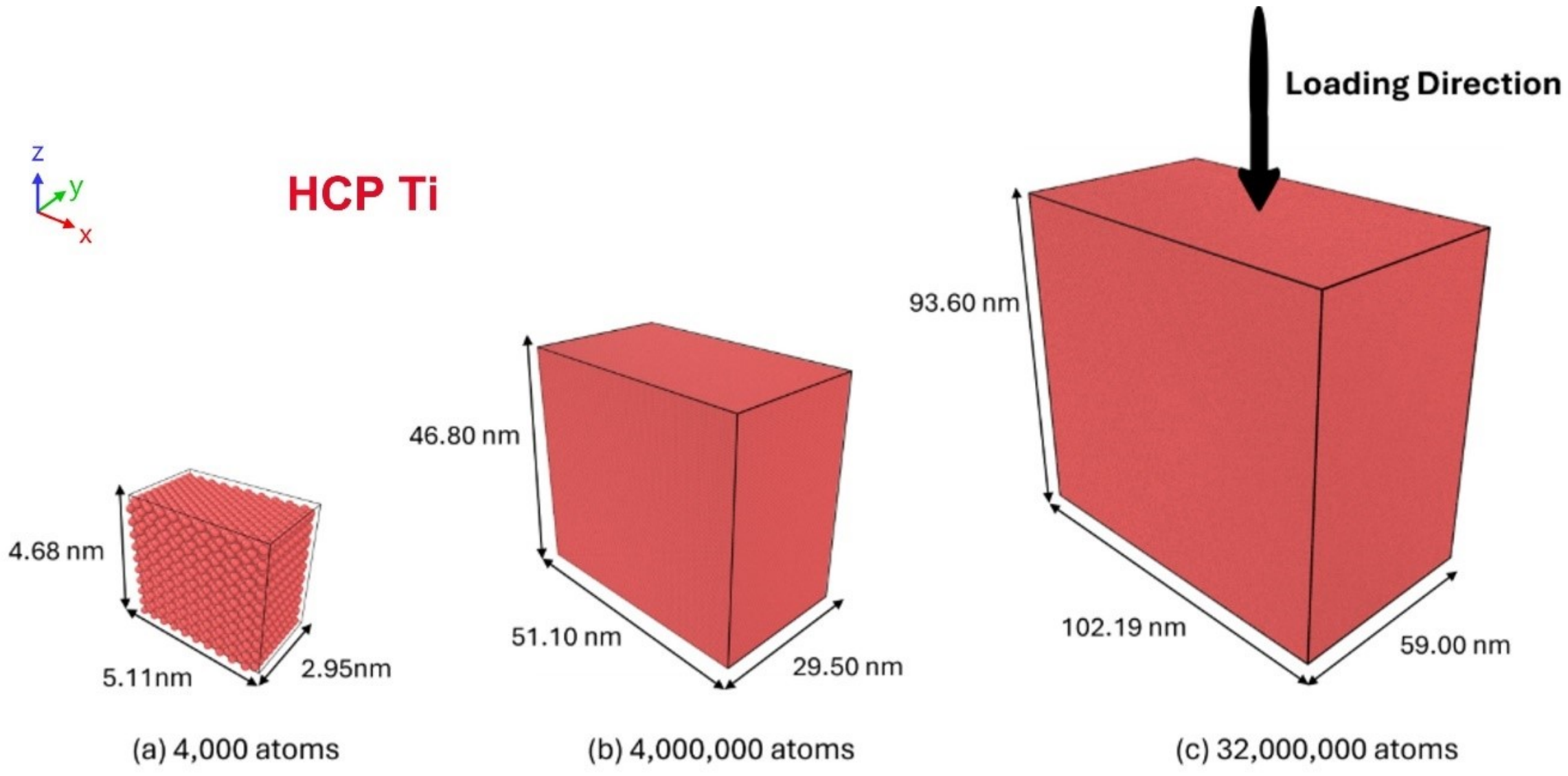

Fig. 1. Three Ti crystalline models of different sizes, shown roughly to scale. From left to right, the system size varies by four orders of magnitude with respect to the total number of atoms, while the extent of the dimensions of the simulation boxes is also highlighted. Ti atoms are shown in red, and the black arrow denotes the loading direction (z-axis compression).

For all the MD simulations, a time-step of 1 fs was used to integrate the equations of motion, while periodic boundary conditions were enforced in all directions (3-dimensional space). The isothermal-isobaric ensemble (constant number of atoms, pressure and temperature, or NPT) was applied to allow the volume of the simulation box to change. The Nosé-Hoover thermostat and barostat were chosen to control the temperature and pressure fluctuations [45]. For each Ti crystalline model, the initial configuration was equilibrated at 300 K and 0 bar with a 100 ps isotropic NPT-MD run, in order to obtain a well-equilibrated structure at these conditions.

Subsequently, uniaxial compression was imposed along the z-axis (c-axis) at a strain rate of $10^9$ $s^{-1}$, while maintaining the temperature at 300 K. Compression was applied using the *fix deform* command as implemented in LAMMPS. The *fix deform* command applies a continuous deformation to the simulation box by adjusting its shape/dimensions at a specified rate. As the box changes shape, atomic positions are remapped to follow the applied strain, allowing the system to respond to the imposed loading. An anisotropic NPT ensemble was employed for the MD simulations during the compression events, where the barostat adjusts the x and y dimensions of the simulation box (target pressure 0 bar), to allow lateral relaxation during loading.

Furthermore, to examine the sensitivity of the system-size effects to the applied loading rate, additional simulations were conducted at strain rates of $5\times10^8$ $s^{-1}$, $10^8$ $s^{-1}$, and $10^{10}$ $s^{-1}$. A total strain of 40% was applied for all strain rates, corresponding to total simulation times of 40, 400, 800, and 4000 ps for the $10^{10}$ $s^{-1}$, $10^9$ $s^{-1}$, $5\times10^8$ $s^{-1}$, and $10^8$ $s^{-1}$ strain rates, respectively. The simulation results, together with the Ti crystalline models before, during and after compression, are visualized with the OVITO (Open Visualization Tool) software [46]. Structural analysis was performed using the Polyhedral Template Matching (PTM) method [47] as well as the coordination-analysis module in OVITO, while dislocations were characterized using the Dislocation Extraction Algorithm (DXA) [48].

# 3. Results and discussion

To ensure that the Ti crystalline structures have been sufficiently equilibrated before compression, NPT-MD runs of 100 ps were performed at 300 K and 0 bar, for all the models. In the Supplementary Information (SI), Figs. S1–S7 show the time evolution of temperature and pressure during the equilibration run for the seven different modelled systems. It can be observed that the temperature fluctuations are quite narrow around the target value of 300 K, while the pressure experiences a larger range of variations, but are still within reasonable limits, around the average value of 0 bar. These highlight that the thermostat and barostat drive the systems consistently to thermodynamic equilibrium, and hence, that all the crystalline structures have been equilibrated before the compression events.

## 3.1. System size effects: $10^9$ $s^{-1}$

The impact of the simulation system size on the deformation behaviour of Ti was first examined at a strain rate of $10^9$ $s^{-1}$, with all seven modelled systems evaluated under identical loading and thermodynamic conditions. The simulation results are discussed in terms of: (a) the stress–strain response, which reflects the macroscopic mechanical behaviour of the atomic structure; (b) the evolution of the crystal structure, which reveals modifications within the HCP lattice and the formation of defects; and (c) the characteristics of the dislocation activity, which identify the underlying atomic-scale mechanisms responsible for plastic flow, and provide insights into the features observed in the relevant stress–strain curves.

### 3.1.1. Stress-strain analysis

In Fig. 2, the stress ($\sigma_{zz}$, i.e., the stress component in the z-direction)-strain ($\varepsilon$) curves are presented for the seven α-Ti modelled systems compressed along the c-axis at a strain rate of $10^9$ $s^{-1}$. All curves display the same linear elastic response up to ~15% strain, with nearly identical slopes. This highlights that the compressive elastic modulus of the Ti models is independent of the simulation system size, a behaviour that is anticipated for a defect-free single crystal under periodic boundary conditions, where the elastic constants are intrinsic properties of the material [49]. In addition, the yield stress, defined as the peak stress, exhibits a similar value for all the modelled systems, reaching about 16.6 GPa at ~15% strain, before a sharp drop occurs. This decrease is indicative of the onset of plastic deformation, and corresponds to the nucleation of the first defect. Hence, the critical resolved shear stress (CRSS) for this nucleation event does not show a dependence on the simulation system size.

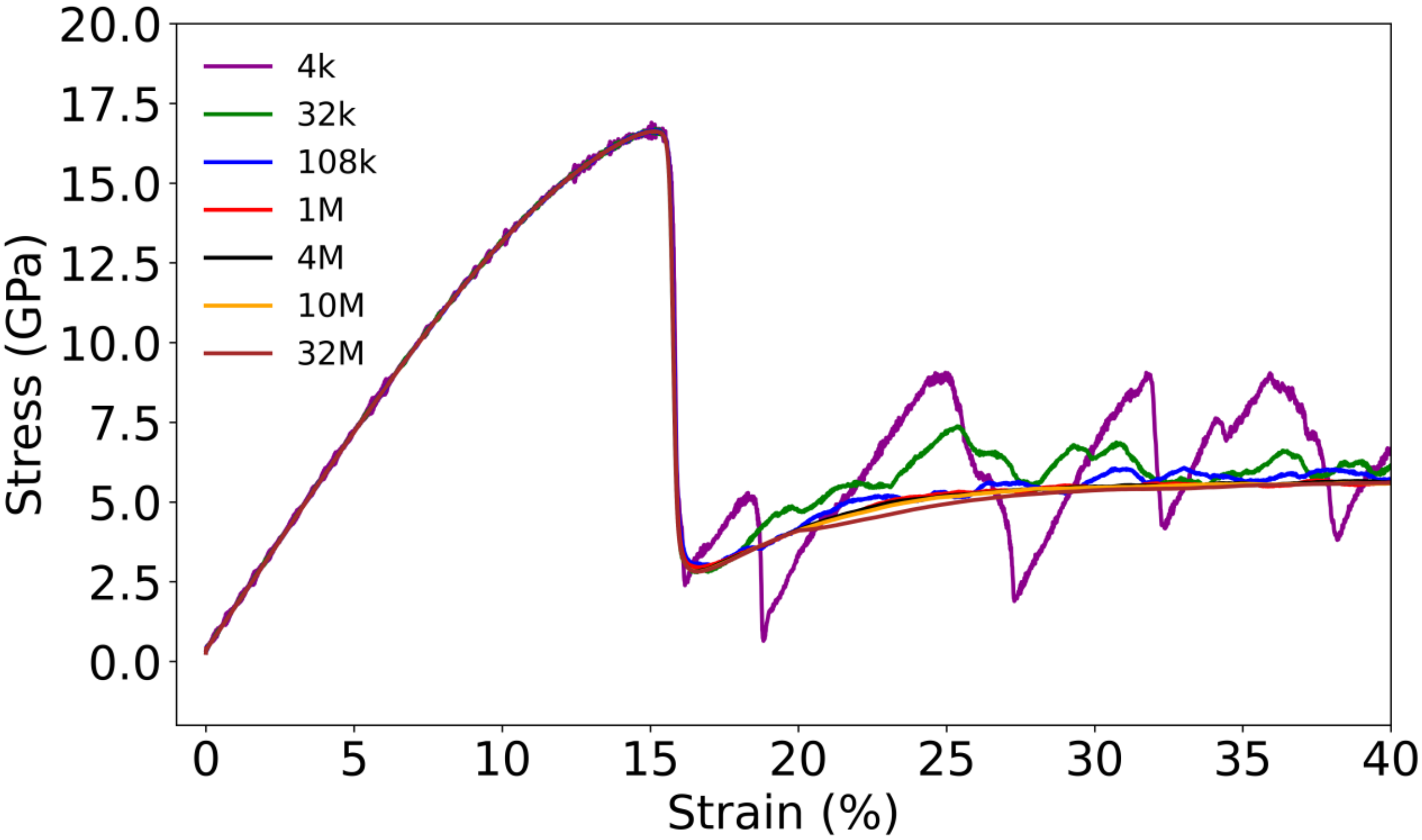

Fig. 2. Stress-strain curves for the seven different Ti modelled structures compressed at a $10^9$ $s^{-1}$ strain rate with molecular-dynamics simulations performed at 300 K.

However, beyond the yield point, clear finite-size effects emerge in the stress-strain curves of the simulated systems, as shown in Fig. 2. The two smallest models (4k and 32k) studied here manifest strong stress oscillations caused by the finite-size limitations. In these crystalline structures, due to the small number of available nucleation sites, a single dislocation or twin can span the entire simulation cell, and a large portion of stored elastic energy can be suddenly released [50]. As the simulation system size increases, the stress response becomes noticeably smoother, and it can be observed that the stress plateaus for the Ti models with 1M atoms and beyond are converged, highlighting no further influence by finite-size effects, at least for the compression events with the $10^9$ $s^{-1}$ applied strain rate.

To further assess the effect of compression, additional metrics are reported in the SI. Figs. S8 and S9 show the evolution of the potential energy per atom and the simulation cell volume, respectively, as a function of time during compression. For the 4k model, both quantities exhibit pronounced fluctuations, indicating strong finite-size effects. The radial distribution functions (RDFs) of the 32k and 32M models before and after compression are shown in Fig. S10. The RDF captures the average atomic structure, with the first peak corresponding to the nearest-neighbour Ti–Ti distance. Compression leads to a shift of this peak to smaller distances (from 2.91 Å to 2.85 Å), and a lowering in peak height. This reduction in the first RDF peak results from shorter Ti–Ti distances under compression and a broader nearest-neighbour distance distribution, which lowers the probability of finding atoms at a single specific distance. Unlike the stress response, no noticeable size effect is observed in the RDFs.

It is noted that, since the 4k model is particularly affected by finite-size artifacts, it is excluded from the subsequent structural analysis presented in this study (although, it is considered in some of the analysis that is provided in the SI).

### 3.1.2. Structural analysis

To track the evolution of the crystal structure during the deformation simulations, PTM was employed using a root-mean-square-deviation (RMSD) cutoff of 0.15. Fig. 3 illustrates the atomistic evolution of the local crystal structures for four selected Ti modelled system sizes (32k, 108k, 4M, and 32M). Atoms identified as HCP, FCC, BCC, and other (or unknown) structures are shown in red, green, blue, and grey, respectively. For each model, four distinct stages during compression are depicted: prior to loading ($\varepsilon = 0\%$), immediately following the onset of plasticity ($\varepsilon = 16.5\%$, corresponding to the first major stress relaxation event), an intermediate stage ($\varepsilon = 25\%$), and the final simulated Ti structure at the end of loading ($\varepsilon = 40\%$).

As shown in Fig. 3, all modelled systems begin as fully HCP at $\varepsilon = 0\%$. After yielding, when the Ti structure enters the plastic deformation regime ($\varepsilon = 16.5\%$, a point that corresponds to the stress drop in Fig. 2), the modifications to the original crystal structure are initiated, and FCC, BCC, and other local atomic environments are formed inside the simulation lattice. In the 32k and 108k models, the transformed regions stay more spatially localized [see Figs. 3(a) and 3(b)], whereas in the 4M and 32M simulated systems the FCC, BCC, and unknown structures exhibit a significantly more uniform distribution within the crystalline lattice [see Figs. 3(c) and 3(d)]. As the applied strain increases further, the smaller simulated systems retain a more heterogeneous deformation pattern, while, in contrast, the larger Ti models develop a homogeneous distribution of the transformed crystal structures. This demonstrates that, as the simulation system size grows, the structural evolution of Ti during compression becomes more uniform, as well as more scattered across the entire simulation cell.

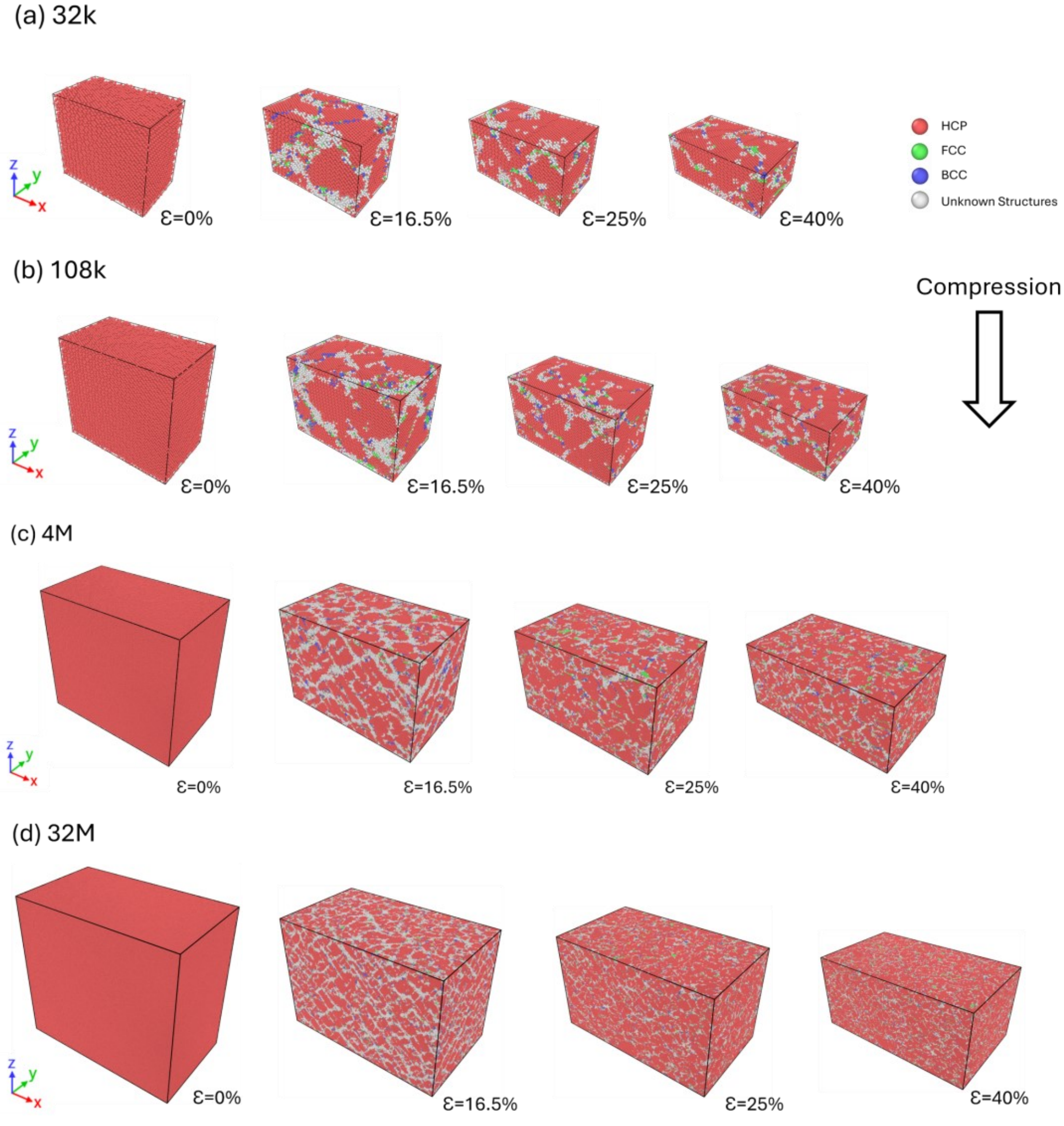

Fig. 3. Dynamic evolution of the different crystal structures in four different Ti models during compression [(a) 32k, (b) 108k, (c) 4M, and (d) 32M], identified using Polyhedral Template Matching (RMSD cutoff = 0.15). Individual atoms are coloured as HCP (red), FCC (green), BCC (blue), and unknown (grey). From left to right, at each panel, snapshots are shown at ε = 0%, 16.5%, 25%, and 40%, and the loading direction is indicated by an arrow (z-axis compression).

Fig. 4 displays the evolution of all the different crystallographic structure fractions as a function of strain for each Ti model, obtained using a PTM analysis. Atoms labelled as “Other” represent unidentified local environments that do not belong to any of the HCP, FCC, or BCC classifications [47]. All simulated structures remain fully HCP throughout the elastic regime, until a sudden decrease in the HCP fraction occurs, as it can be seen in Fig. 4(a). This drop coincides with a rapid increase in FCC, BCC, and other structures, shown in Figs. 4(b-d), in consistency with the yield point and stress drop observed in Fig. 2, and it is indicative of the onset of the plastic deformation of the crystal structure. The strain value at which this transition occurs is nearly identical for all the Ti models, ranging between 15.11% and 15.22%. The sharp reduction in the HCP fraction takes place over a narrow window during the loading event (with a width of ~1.5% after the initial point of the drop), and all the simulated systems exhibit a partial recovery of the HCP structure. The minimum HCP amounts, before recovery, are 64.7%, 68.5%, 66.2%, 65.4%, 65.9%, and 66.1% for the 32k, 108k, 1M, 4M, 10M, and 32M models, respectively.

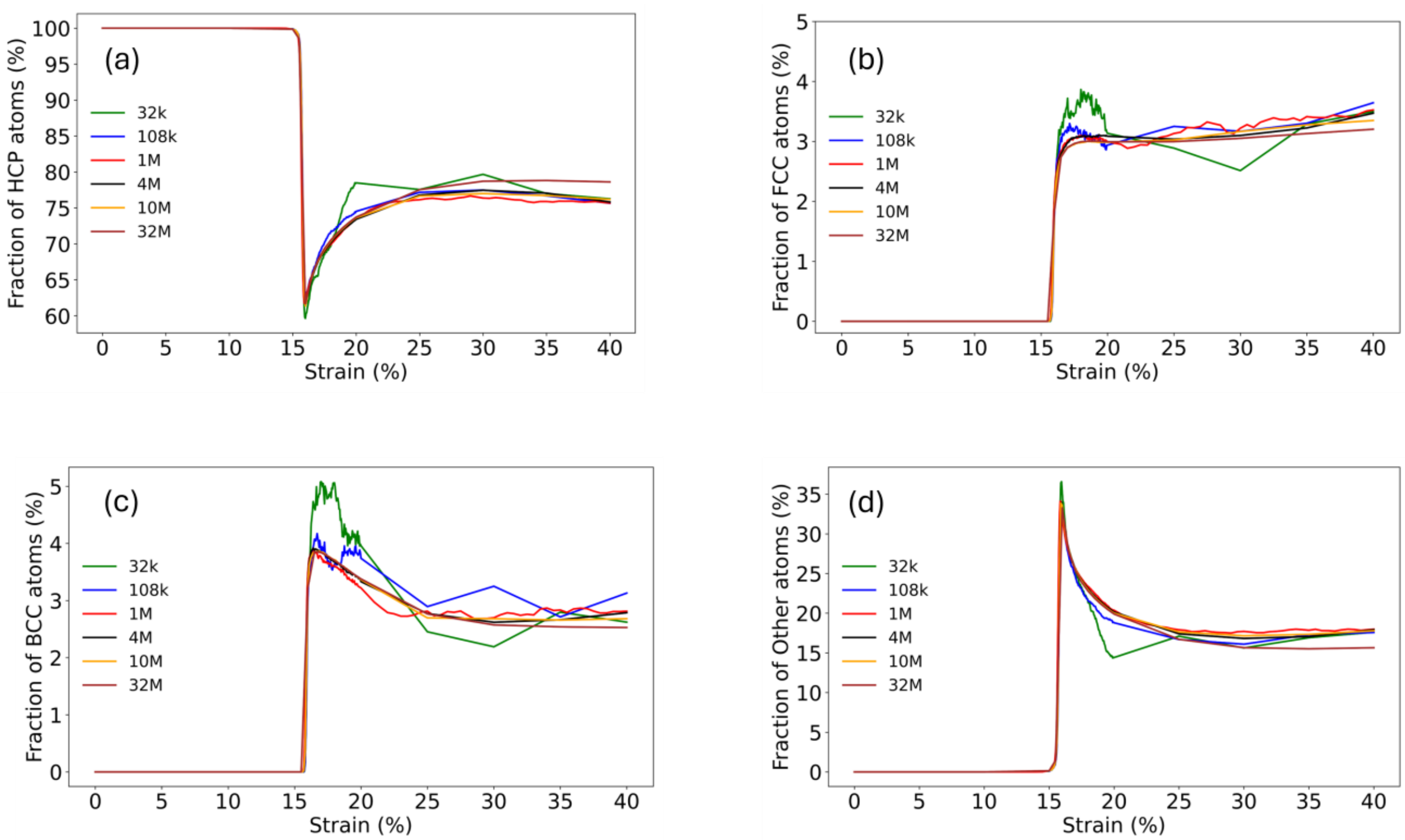

Fig. 4. Dynamic evolution of crystallographic structure fractions as a function of strain for Ti models of different system sizes during the compression simulations with a $10^9$ $s^{-1}$ strain rate. The proportions of HCP (a), FCC (b), BCC (c), and Other (d) structures are shown for each model.

In the plastic region, similarly with the observations from the stress-strain results in Fig. 2, the temporal behaviour of the crystallographic structure fractions exhibits a dependence on the simulated system size. For the smaller Ti models (namely, 32k, 108k, and 1M), significant fluctuations in all structure types are detected, whereas the three largest modelled systems (≥4M) display a more monotonic evolution, with a gradual development of the different atomic structures inside the simulated systems during the compression event, and a faster equilibration. For instance, in the 32k model, the HCP ratio is 71.7% at the strain value of 15.8%, accompanied by 28.0% of other structures, while at 20% strain, the HCP amount increases to 75.6% and the other fraction decreases to 17.5%. In contrast, the modifications in the largest system size studied here (32M) show a more moderate rate of change: at 15.8% strain, the HCP percentage is 73.7% along with 25.7% other structures, and at 20% strain the HCP fraction is 72.8%, while the other structures comprise 20.7%. Remarkably, the 32M Ti model converges to a higher proportion of HCP atoms (correspondingly to a lower amount of other atoms) compared to any other simulated system considered here, indicating that increasing system size leads to potentially a different prediction of the crystal structural evolution during plastic deformation.

Although the percentages of HCP and other structures are found to converge for models larger than 32k atoms, the same level of convergence is not demonstrated for the fractions of the FCC and BCC atoms inside the different Ti models, as it can be seen in Figs. 4(b) and 4(c). Even in the 1M-atom model, the FCC and BCC contents continue to fluctuate at the end of the compression simulation. This variability arises from the relatively low amount of these crystalline environments in the simulated structures, compared to HCP and other, which makes them more sensitive to statistical fluctuations. Consequently, larger simulation systems are required to reliably represent the evolution of the small quantities of these structural components during the compression events.

### 3.1.3. Dislocation analysis

A dislocation analysis is essential for investigating the atomic-scale mechanisms governing plastic deformation, since the nucleation and propagation of specific dislocations control the plastic response of crystalline materials, and hence directly influence their mechanical behaviour [51]. For Ti compressed along the c-axis, the deformation is primarily accommodated by prismatic ⟨a⟩ slip, which exhibits the lowest CRSS at room temperature [52,53], and by pyramidal ⟨c+a⟩ slip, which is associated with significantly higher activation stresses, and is therefore less frequently activated under the same conditions [54,55]. To characterise the evolution of the dislocation activity during the compression simulations in this study, dislocations were identified using the Dislocation Extraction Algorithm (DXA) implemented in OVITO [48].

Fig. 5 shows the distributions of all the different dislocation types calculated at the end of the compression simulations (i.e., 40% strain, $10^9$ $s^{-1}$ strain rate) for three of the modelled systems studied here (108k, 1M, and 4M). The analysis reveals the presence of distinct dislocation types, based on their unique Burgers vectors, in the simulated Ti structures, including $\frac{1}{3} < 1\bar{2}10 >$ prismatic dislocations, $\frac{1}{3} < 1\bar{1}00 >$ partial dislocations on basal plane, and $\frac{1}{3} < 1\bar{2}13 >$ pyramidal dislocations. Such dislocations correspond to the main slip mechanisms activated during the deformation of Ti. Some of the dislocations that were formed inside the simulated structures could not be identified by the DXA analysis, and therefore are classified as "Other" types of dislocations.

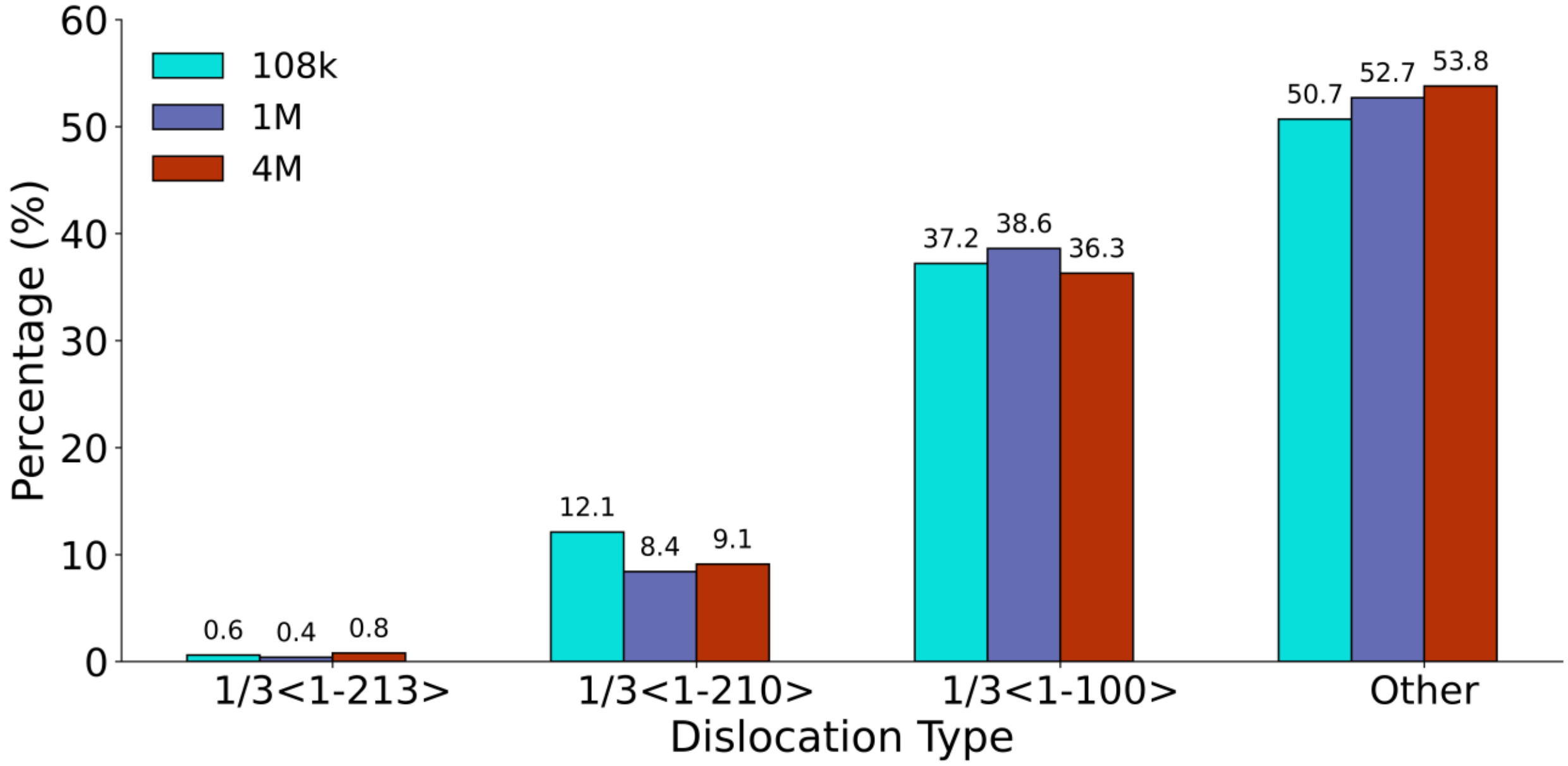

Fig. 5. Contribution (in %) of the different dislocation types identified by DXA for three Ti model sizes (simulated systems 108k, 1M, and 4M) at the end of the $10^9$ $s^{-1}$ compression simulation.

It can be observed that the other dislocations correspond to the largest proportion in all three models, and increase slightly with the simulated system size, measured 50.7%, 52.7%, and 53.8% in the 108k, 1M, and 4M modelled systems, respectively. The $\frac{1}{3} < 1\bar{1}00 >$ partial dislocations prevail as the dominant well-defined dislocation type, calculated approximately 36-39% across all models, with no clear dependence on the system size. The percentage of the $\frac{1}{3} < 1\bar{2}10 >$ prismatic dislocations decrease as the simulated system size increases; from 12.1% in the 108k model to less than 10% in both the 1M and 4M models. The amount of the pyramidal $\frac{1}{3} < 1\bar{2}13 >$ dislocations is very small in all three models (<1%), due to their high activation stresses under the loading conditions investigated in this study. The calculated results suggest that increasing the simulated system size does not activate any new slip paths, but instead affects the relative distribution of the different dislocation types identified in the Ti models, leading to a smaller amount of prismatic slip and a larger quantity of complex dislocation structures.

The time evolution of the total dislocation density and the contributions of the different dislocation types in the three system sizes (108k, 1M, and 4M) are presented in Fig. 6. Following the same trend as in Fig. 5, dislocations classified as other have the largest contribution to the total dislocation density in all three models, while the fractions of the basal $\frac{1}{3} < 1\bar{1}00 >$ partial dislocations and the $\frac{1}{3} < 1\bar{2}10 >$ prismatic dislocations indicate a smaller contribution. The pyramidal $\frac{1}{3} < 1\bar{2}13 >$ dislocations remain negligible throughout the deformation process of the simulated structures.

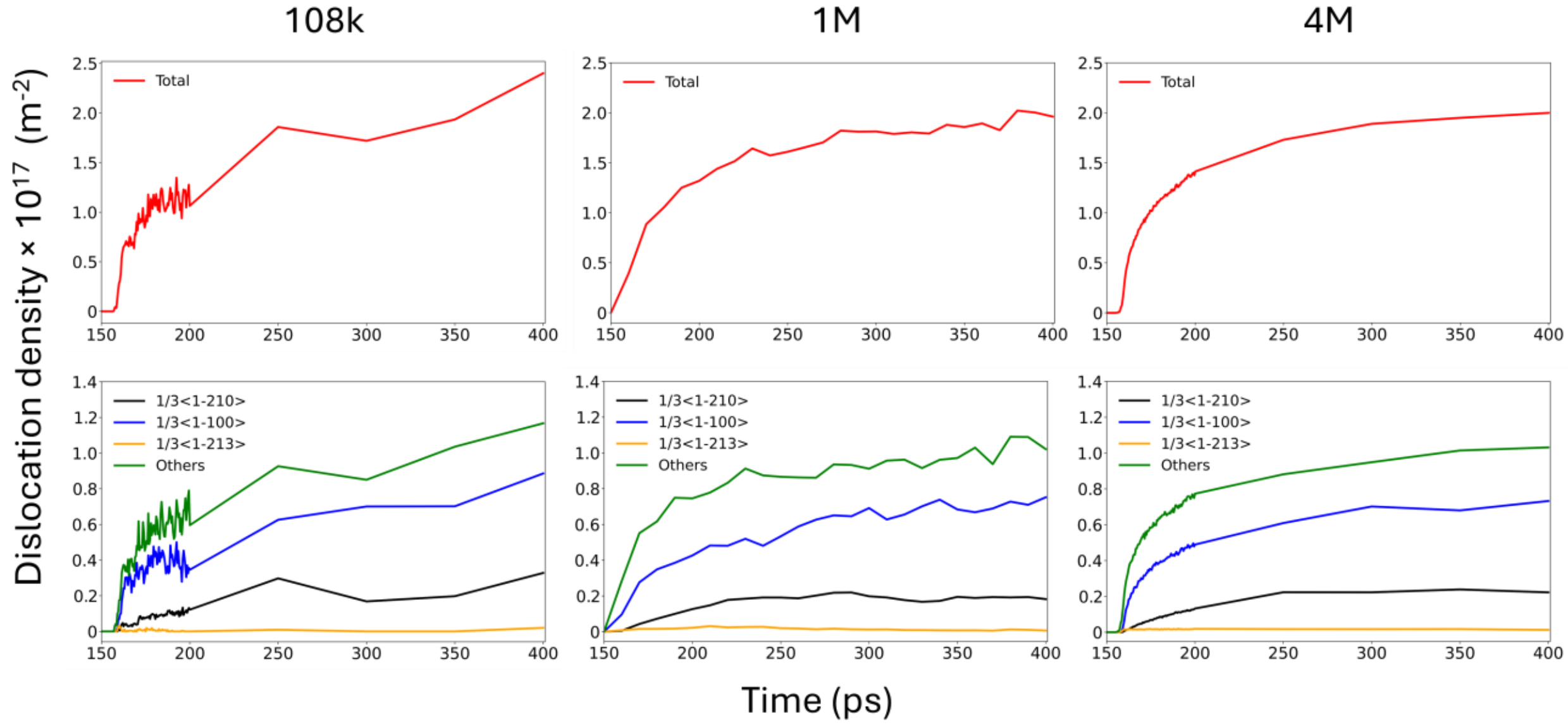

Fig. 6. Time evolution of the total (upper panels) and individual types (lower panels) dislocation densities calculated with DXA during the $10^9$ $s^{-1}$ compression simulation of three different Ti model structures (108k, 1M, and 4M atoms).

A clear dependence on the simulated system size can be observed in the dynamical behaviour of the dislocation density. The 108k Ti model exhibits significant fluctuations in the early stages of the deformation event, as well as higher values for all the dislocation types throughout the simulation compared to the two other larger models. For this model, at 400 ps (corresponding to 40% strain), the total dislocation density reaches approximately $2.4 \times 10^{17}$ $m^{-2}$, which is noticeably higher than the values obtained for the larger Ti models ($2.0 \times 10^{17}$ $m^{-2}$), suggesting that a 108k-atom structure is insufficient to capture the characteristic length scales of dislocation interactions in crystalline Ti [53]. In addition, the predicted value of dislocation density continues to increase towards the end of the MD simulation (as shown in the upper-left panel of Fig. 6), revealing that the deformation response in this system has not been fully stabilised within the simulated time.

For the 1M simulated structure, the overall trend becomes steadier compared to the 108k model, but still exhibits apparent fluctuations in the evolution of the dislocation density, especially for the other dislocations. The total dislocation density reaches a value of $2.0\times10^{17}$ $m^{-2}$ at 400 ps, but the divergence does not allow to obtain a stable behaviour at the end of the compression simulation. In contrast, the 4M Ti model shows a well-defined converged evolution for both the total and the individual dislocation densities. After the initial near-exponential increase, corresponding to the onset of plastic deformation, the dislocation densities evolve more gradually and approach a plateau value ($2.0\times10^{17}$ $m^{-2}$). This behaviour suggests that larger simulation cells provide a more stable representation of the dislocation density evolution during compression, while the smaller models tend to manifest larger fluctuations, and potentially over-estimate the total dislocation density in the crystalline modelled structure.

## 3.2. System size effects: multiple strain rates

In order to study the effect of the strain rate on the deformation behaviour of crystalline Ti, compression simulations were performed at three additional strain rates of $10^{10}$ $s^{-1}$, $5\times10^{8}$ $s^{-1}$, and $10^{8}$ $s^{-1}$ for three selected system sizes from the ensemble of the studied Ti models, namely 108k, 1M, and 4M atoms. All MD simulations were carried out up to a total strain of 40% and at a temperature of 300 K, with the simulation time adjusted according to the applied strain rate (40 ps for $10^{10}$ $s^{-1}$, 800 ps for $5\times10^{8}$ $s^{-1}$, and 4000 ps for $10^{8}$ $s^{-1}$).

Fig. 7(a) presents the stress-strain curves resulted from the simulations with the strain rate of $10^{8}$ $s^{-1}$ for the different modelled systems. All structures exhibit the same elastic response before the yield point (in agreement with the observations from Fig. 2 and the simulation with the $10^{9}$ $s^{-1}$ strain rate), while the yield strain shifts slightly to a lower value as the simulation box size increases. However, the models show a different response after the onset of the plastic deformation, with fluctuations becoming more prominent for all the simulated systems. Notably, the 1M and 4M models exhibit fluctuations even at 40% strain, whereas at $10^{9}$ $s^{-1}$ (see Fig. 2) the same models showed a more stable behaviour with a tendency to reach a plateau at the end of the equivalent simulation with respect to the loading (total strain). Such response suggests that at lower strain rates, even the 4M model cannot be assumed to be fully representative in describing the deformation behaviour of the compressed Ti structure.

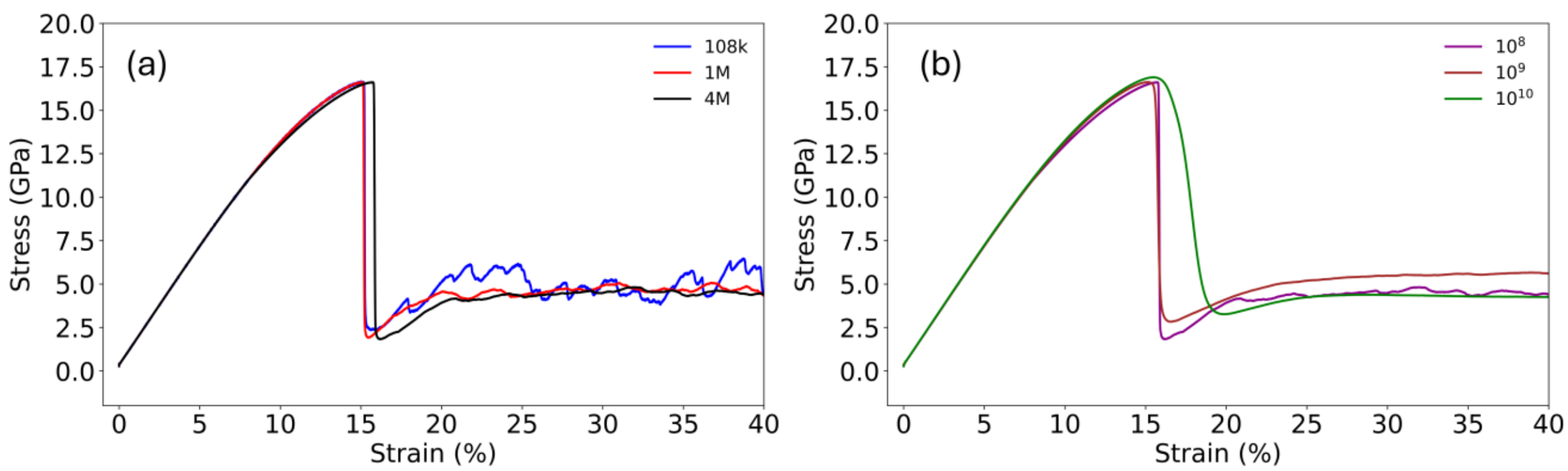

Fig. 7. Stress-strain analysis: (a) at a strain rate of $10^8$ s$^{-1}$ for three different Ti modelled system sizes (108k, 1M, and 4M); and (b) for the 4M model at three different strain rates ($10^8$ s$^{-1}$, $10^9$ s$^{-1}$, and $10^{10}$ s$^{-1}$).

Fig. 7(b) shows the stress-strain curves for the 4M model at three different strain rates, varying by one order of magnitude ($10^{10}$, $10^9$, and $10^8$ s$^{-1}$). A similar trend is observed in the elastic region for all strain rates, while the yield stress at $10^{10}$ s$^{-1}$ is slightly higher compared to that obtained at the two other strain rates applied here. This can be attributed to the limited time that is available for dislocation nucleation, due to the fast compression during the $10^{10}$ s$^{-1}$ simulation. This results in a strain-rate strengthening effect where a higher stress is required to initiate plastic deformation [56–58]. It can be observed that after the onset of plastic deformation, the compression events with strain rates $10^9$ s$^{-1}$ and $10^8$ s$^{-1}$ show a sharper stress drop, indicating a more sudden stress relaxation, compared to the simulation with the $10^{10}$ s$^{-1}$ strain rate. This implies that at higher strain rates the plastic relaxation is kinetically limited and dislocations cannot accommodate the imposed deformation instantaneously, which leads to a more gradual decrease in stress over a wider strain range.

After reaching the minimum, the equilibration of the stress curves is taking place at different stress levels for the different strain rates showing a different behaviour for the simulated Ti model. Typically, at very high strain rates the rapid deformation increases the structural disorder, which, in turn, reduces the flow stress. In contrast, at lower strain rates the dislocations have sufficient time to nucleate and propagate, allowing stress relaxation, leading to more fluctuations in the stress-strain response. These results suggest that system size and strain rate are correlated and should be considered jointly when studying the deformation behaviour of a crystalline material. The same analysis for the 108k and 1M modelled systems is provided in the SI (Fig. S11).

Fig. 8 presents the distributions of the various crystallographic structures (identified using a PTM analysis) for the 4M model at 40% strain for three different strain rates ($10^8$, $5\times10^8$, and $10^9$ s$^{-1}$). As illustrated by the histograms, HCP is the dominant structure inside the Ti model, followed by the “Other” structures for all the strain rates considered here. In the deformation simulation with the lowest strain rate ($10^8$ s$^{-1}$) the proportion of the HCP fraction has the highest value (89.3%), which then decreases with increasing strain rate to 81.7% for 5×$10^8$ $s^{-1}$ and 75.8% for $10^9$ $s^{-1}$. In contrast, the total amount of other structures increases with increasing strain rate, with the lowest value being 7.5% at $10^8$ $s^{-1}$. Also, a similar behaviour can be observed for the FCC and BCC fractions, which remain at relatively low percentages of 1.8% and 1.4%, respectively, for the $10^8$ $s^{-1}$ strain rate.

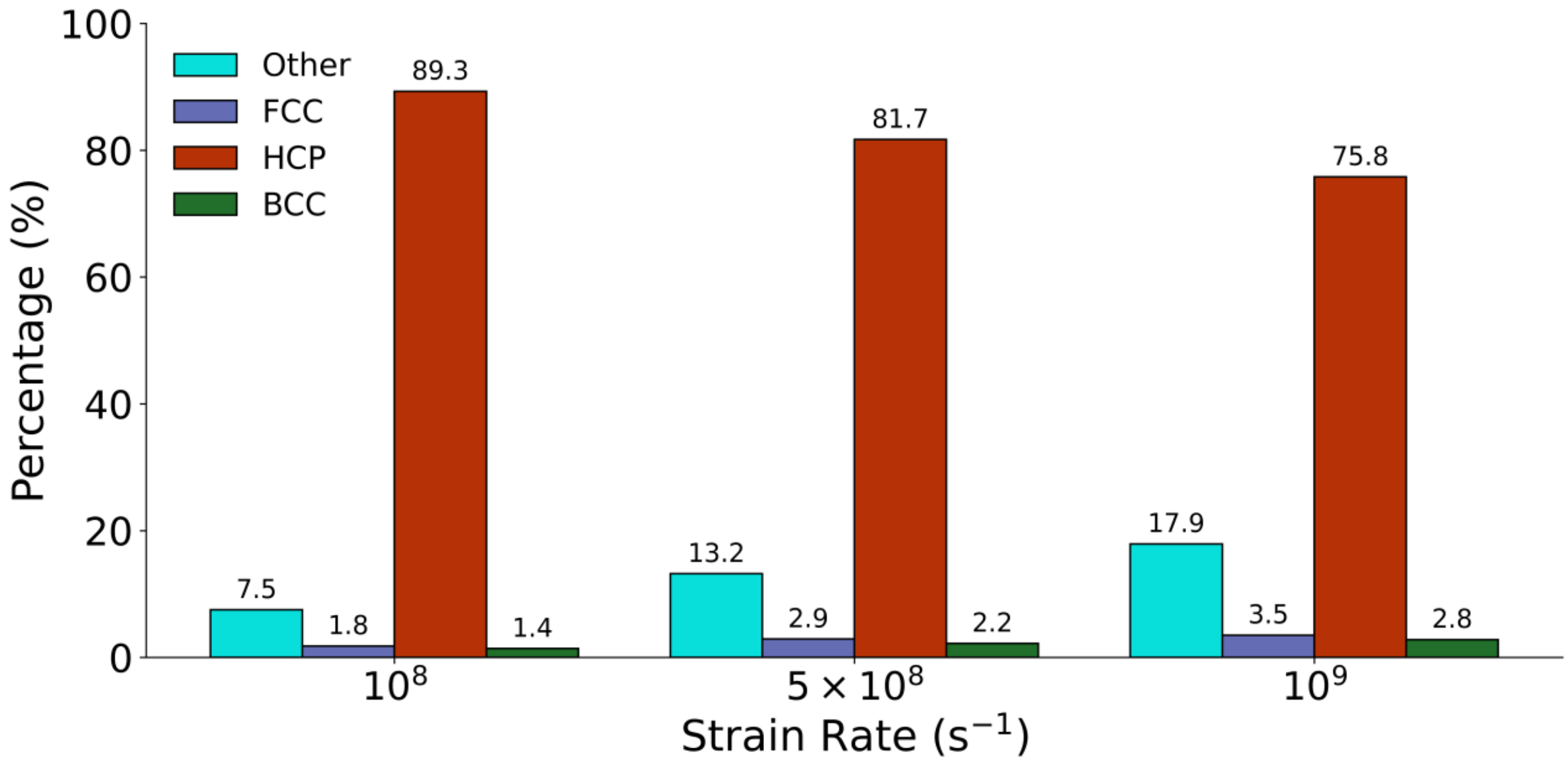

Fig. 8. Distributions of crystallographic structures for the 4M-atom Ti model at 40% strain for three different applied strain rates ($10^8$, 5×$10^8$, and $10^9$ $s^{-1}$), obtained using a Polyhedral Template Matching analysis. The fractions of HCP, FCC, BCC, and "Other" structures are shown for each strain rate.

Based on the analysis presented in Fig. 8, at a strain rate of $10^8$ $s^{-1}$ the simulated system retains most of its initial structure (pure HCP), and minor phase transformations occur inside the 4M crystalline model. The evolution of the structural transformations for the deformation events at this strain rate in three different modelled structures was further examined in order to clarify the atomic-scale modifications inside the simulated Ti systems. Fig. 9 presents the evolution of the crystallographic structure fractions as a function of strain for Ti models of different system sizes (108k, 1M, and 4M) at the strain rate of $10^8$ $s^{-1}$. All the models follow a similar trend in the elastic region up to the yield point (~15-16% strain), in accordance with the behaviour observed from the stress-strain analysis [Fig. 7(a)]. The abrupt decrease in Fig. 9(a) corresponds to the phase transformation from HCP to BCC, FCC and "Other" structures (onset of plastic deformation), which shows a slight variation among the different simulated system sizes. In the plastic region, a clear size effect can be observed for all the structure types. For instance, the minimum HCP fraction has the lowest amount for the 1M Ti model, reaching 72.8% at 15.3% strain, while the corresponding values are 80.1% at 15.5% strain and 78.4% at 16% strain for the 108k and 4M simulated structures, respectively.

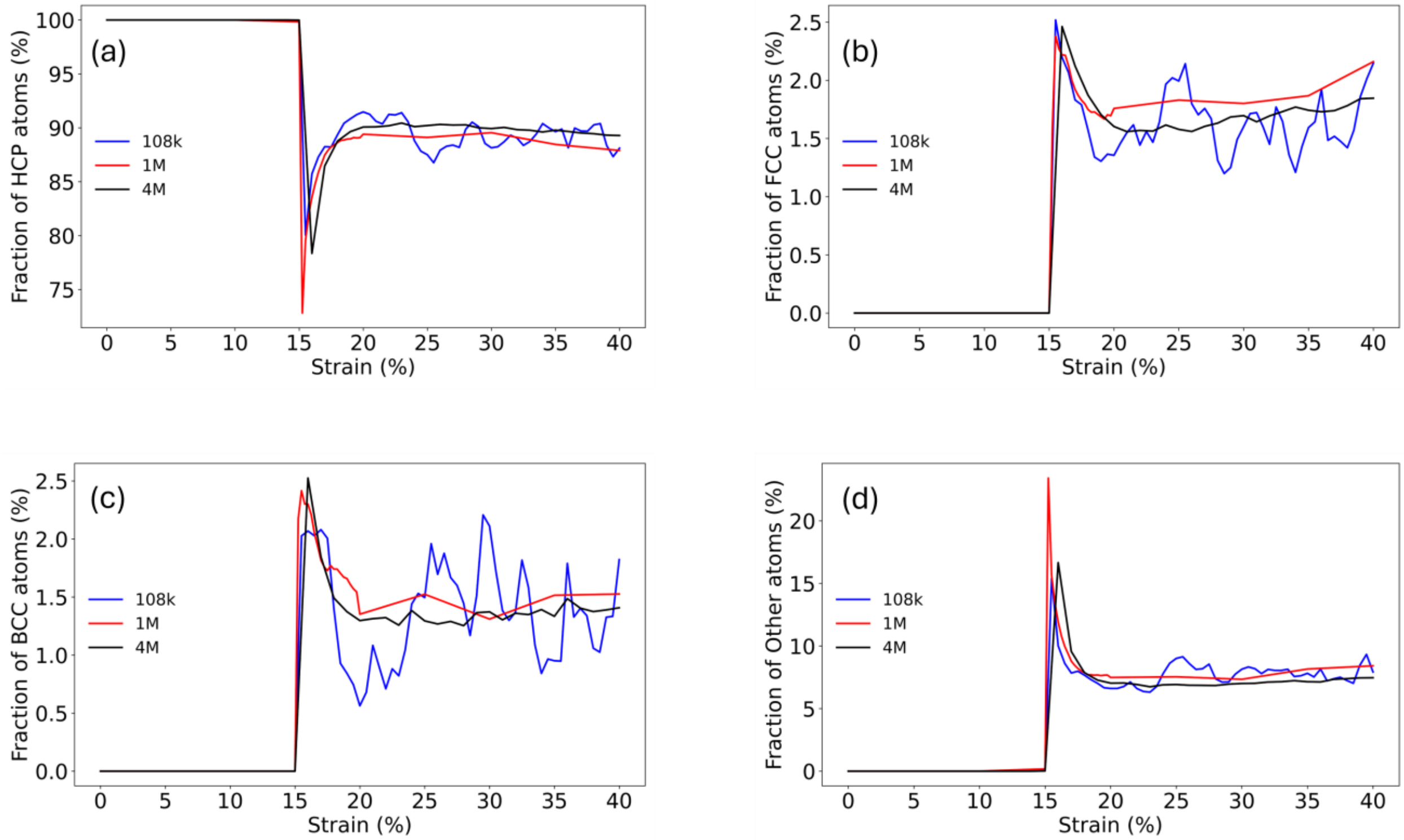

Fig. 9. Dynamics of crystallographic structure fractions as a function of strain during the compression simulations of Ti models of three different system sizes (108k, 1M, and 4M atoms) at a $10^8$ $s^{-1}$ strain rate. The percentages of HCP (a), FCC (b), BCC (c), and Other (d) structures are shown for each model.

All the models show a partial recovery of the HCP structure, with the 108k and 1M simulated systems exhibiting more fluctuations compared to the 4M Ti model. However, it can be seen that even the largest modelled structure (4M) used in this set of deformation events does not fully stabilize. Such behaviour makes the validation of the equilibration of the deformation simulation a challenging task. A similar pattern is observed in Figs. 9(b-d), where all the different obtained structure types follow a trend opposite to that of the HCP fraction. These results indicate that at low strain rates larger simulation cells are required to achieve optimal convergence, and to describe the deformation behaviour of the crystalline structure in a sufficient way.

Considering that the evolution of the atomic structure is associated with the nucleation of defects and dislocations inside the simulated system, a dislocation density analysis was performed for the different strain rates. Fig. 10 displays the evolution of the total dislocation density for the 4M Ti model at three applied strain rates, specifically, $10^8$, $10^9$, and $10^{10}$ $s^{-1}$. For the highest value of the strain rates employed here ($10^{10}$ $s^{-1}$), the predicted dislocation density is very low. This can be explained due to the rapid deformation, which limits the time available for defect formation within the modelled structure. At the end of the compression simulation (40% strain), the total dislocation density for the $10^{10}$ $s^{-1}$ strain rate is $4.5\times10^{15}$ $m^{-2}$, which differs by two orders of magnitude from the corresponding value obtained from the deformation simulation with the $10^9$ $s^{-1}$ strain rate.

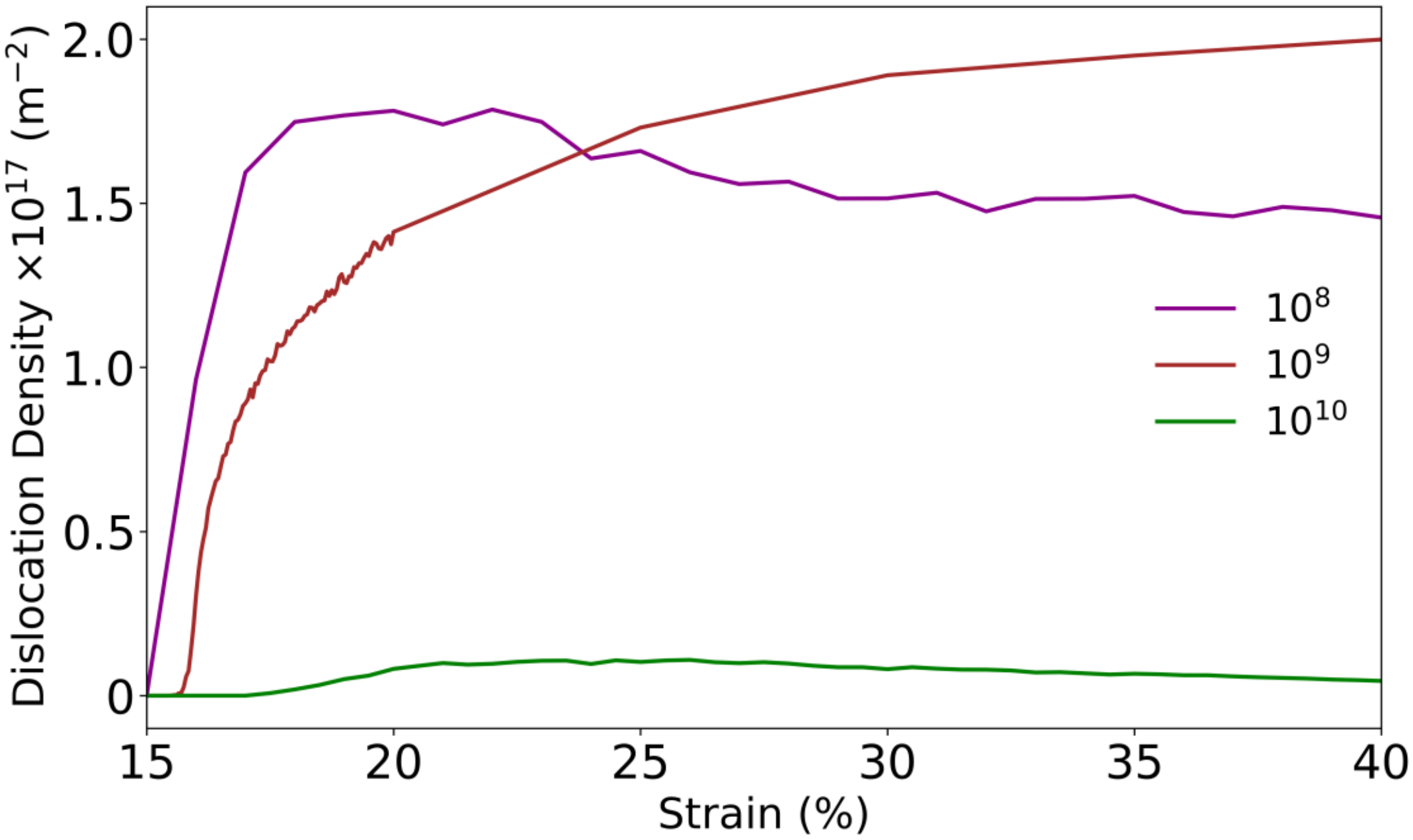

Fig. 10. Evolution of the total dislocation density for the 4M-atom Ti model as a function of strain (%), at three different applied strain rates ($10^8$, $10^9$, and $10^{10}$ $s^{-1}$).

During the compression simulations with the $10^9$ and $10^8$ $s^{-1}$ strain rates a significant increase in the dislocation density after the onset of the plastic deformation is observed in the Ti modelled structure (Fig. 10). This increase is slightly delayed in the compression event with the $10^9$ $s^{-1}$ strain rate, a finding that is associated with the small difference in the yield strain identified in the respective stress-strain curves [Fig. 7(b)]. In addition, the evolution of the dislocation density exhibits a different behaviour between these two strain rates. For the case of the $10^9$ $s^{-1}$ a near-exponential increase of the dislocation density is evident during the deformation event, whereas for the $10^8$ $s^{-1}$ a sharp increase is followed by a decrease of the predicted dislocation density during the simulation. At the end of the deformation, a higher and more stable value (~$2.0\times10^{17}$ $m^{-2}$) is reached in the simulation with the $10^9$ $s^{-1}$ strain rate, while in the respective process with the $10^8$ $s^{-1}$ strain rate more fluctuations are apparent, and a lower predicted value (~$1.5\times10^{17}$ $m^{-2}$) is obtained for the total dislocation density. These results indicate that as the applied strain rate is getting lower, larger simulated systems are required to achieve a stable dislocation evolution and a well equilibrated final estimated value for the dislocation density, while even relatively large models (such as 4M atoms) may be inadequate for detecting the deformation activity of the crystalline material.

Fig. 11 shows the evolution of dislocation density for each dislocation type at the three different strain rates ($10^8$, $10^9$, and $10^{10}$ $s^{-1}$) for the 4M Ti model. For all the applied strain rates the same dislocation types are found inside the modelled system, with different relative amounts. Also, it can be observed that the "Other" dislocations account for the largest proportion of the total dislocation density, with values that are significantly higher (by approximately one order of magnitude) than those of the rest of the identified dislocation types (as reflected by the y-axis scale). In accordance with the observations from Figs. 5 and 6, the $\frac{1}{3} < 1\bar{1}00 >$ dislocations constitute the second largest proportion, followed by the $\frac{1}{3} < 1\bar{2}10 >$ dislocations, while the amount of the $\frac{1}{3} < 1\bar{2}13 >$ dislocation type remains negligible. It is noted that the pattern observed in Fig. 11(d) closely follows that of the total dislocation density (Fig. 10), due to the dominant contribution of the other type of dislocations inside the modelled crystalline structure.

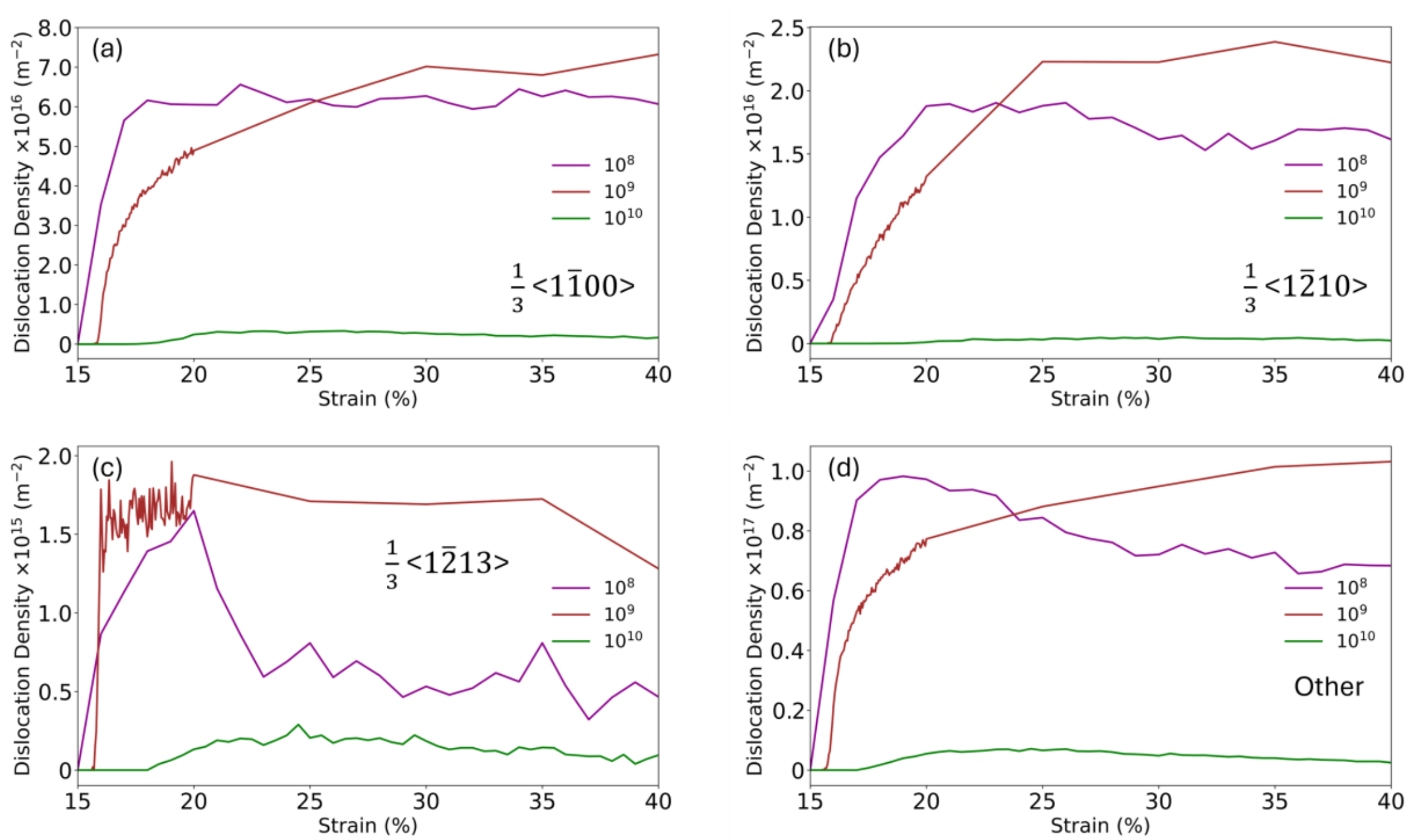

Fig. 11. Temporal behaviour of different dislocation types at different strain rates ($10^8$, $10^9$, and $10^{10}$ $s^{-1}$) for the 4M Ti model: (a) basal $\frac{1}{3} < 1\bar{1}00 >$, (b) prismatic $\frac{1}{3} < 1\bar{2}10 >$, (c) pyramidal $\frac{1}{3} < 1\bar{2}13 >$, and (d) Other dislocations.

Furthermore, the overall trends of the individual dislocation types are consistent with those observed for the total dislocation density in Fig. 10. For the simulation with the $10^{10}$ $s^{-1}$ strain rate the high imposed compression rate restricts the time available for dislocation nucleation leading to an insufficient description of the deformation characteristics. In contrast, in the simulations with the $10^9$ and $10^8$ $s^{-1}$ strain rates the dislocations exhibit a pronounced dynamical progression over the applied strain. A near-exponential behaviour is observed for $10^9$ $s^{-1}$ for some of the dislocation types in the Ti model, with a tendency for equilibration during the MD simulation [Figs. 11(a), (b), and (d)]. For $10^8$ $s^{-1}$ the dislocation evolution exhibits more fluctuations, indicating a less stable dislocation activity during the compression simulation. These results highlight that, while the applied strain rate does not significantly affect the relative ratios of each dislocation type, it has an impact on the dynamical response of the dislocations during the deformation process in crystalline Ti.

# 4. Conclusion

Molecular-dynamics simulations were employed to investigate the deformation behaviour of α-Ti under compression. We have explored the effects of model size, varied by four orders of magnitude, up to 32 million atoms, and of strain rate, varied by two orders of magnitude, down to $10^8$ $s^{-1}$, on the mechanical response, structural evolution, and dislocation dynamics of the simulated systems.

The calculations show that the intrinsic material properties, such as the elastic response and yield point, are independent of both system size and strain rate. However, finite-size effects emerge at the onset of plastic deformation, associated with defect nucleation, a sharp decrease in the HCP fraction and a simultaneous phase transformation to FCC, BCC, and "other" crystallographic structures, within a narrow strain window. In the plastic regime, models of smaller size exhibit pronounced stress oscillations, whereas larger simulation cells show a more stable dynamical response and a clear equilibration trend. At the end of the compression events, the small modelled systems display a heterogeneous distribution of structural environments, while the larger models have a more homogeneous deformation response.

The results of the dislocation analysis highlight the dependence of the deformation behaviour of the simulated system on model size. Small modelled structures show a continuous increase in dislocation density with significant fluctuations, whereas large Ti models exhibit a near-exponential increase followed by a plateau, indicating a stabilised dislocation activity. The dominant deformation mechanisms remain the same in all studied models and deformation scenarios, suggesting that varying system size and strain rate does not trigger any new slip systems, but instead impact their dynamical evolution and relative activity inside the crystalline structure.

It was found that as the applied strain rate decreases towards more experimentally relevant values ($10^8$ $s^{-1}$), even relatively large modelled systems (4M atoms) do not fully equilibrate within the corresponding simulated time. This indicates that larger system sizes than those typically employed in compression simulations of Ti are required to reliably describe the deformation behaviour of the alloys at low strain rates. We would like to note that identifying the critical minimum simulated system size needed for convergence at different strain rates, and especially at very low rates close to experimental conditions, corresponds to an arduous task for MD simulations. Alternative computational approaches, such as coarse-grained methods, could be explored to address the limitations in length and time scales, with the caveat that the atomic-scale deformation processes may not be accurately represented under these formulations.

Understanding the deformation behaviour of Ti alloys under compression is imperative with respect to their application in complex engineering processes, such as solid-state hydrogen storage, for example. The calculations presented in this study provide an atomistic description of structural and mechanical aspects concerning size effects. The results highlight that system size and strain rate are correlated and should not be considered independently, as their combined effect determines the simulated deformation behaviour of the crystalline structure.

# CRediT authorship contribution statement

**Fatemeh Safari:** Writing – original draft, Investigation, Methodology, Formal analysis, Data curation, Visualization, Software, Validation, Conceptualization.

**Konstantinos Konstantinou:** Writing – review & editing, Supervision, Investigation, Methodology, Conceptualization, Formal analysis, Data curation, Validation, Resources, Funding acquisition.

# Declaration of competing interest

The authors declare that they have no known competing financial interests or personal relationships that could have appeared to influence the work reported in this paper.

# Acknowledgements

F.S. and K.K. acknowledge financial support from Profi 7: Strengthening the Profiling of Universities under joint funding decision No. 352727. K.K. acknowledges financial support from the Research Council of Finland under grant No. 364241 ("NoneqRSMSD"). The authors wish to acknowledge the CSC–IT Center for Science, Finland, for computational resources. We would like to thank Prof A. Ganvir and Dr V. Gidla for useful discussions.

# Data availability

Data supporting the findings of this study will be made publicly available via Zenodo upon publication of the article.

# Uniaxial compression of crystalline HCP titanium: an atomistic modelling study of size effects

Fatemeh Safari [a,*], Konstantinos Konstantinou [a,*]

[a] Department of Mechanical and Materials Engineering, University of Turku, Vesilinnantie 5, 20500 Turku, Finland

[*]Corresponding authors:

konstantinos.konstantinou@utu.fi

fatemeh.f.safari@utu.fi

## Supporting Information

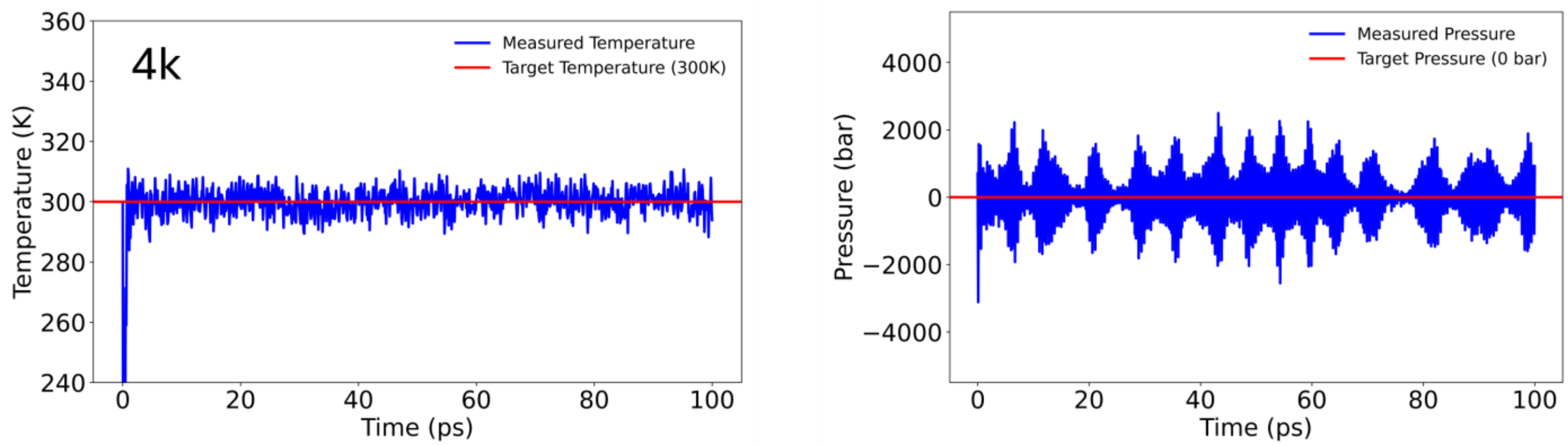

Fig. S1. Time evolution of temperature (left) and pressure (right) during the molecular-dynamics simulation of the equilibration stage for the 4k-atom Ti model. Both quantities fluctuate around the target values of 300 K and 0 bar, indicating that thermodynamic equilibration is achieved prior to compression for the simulated system.

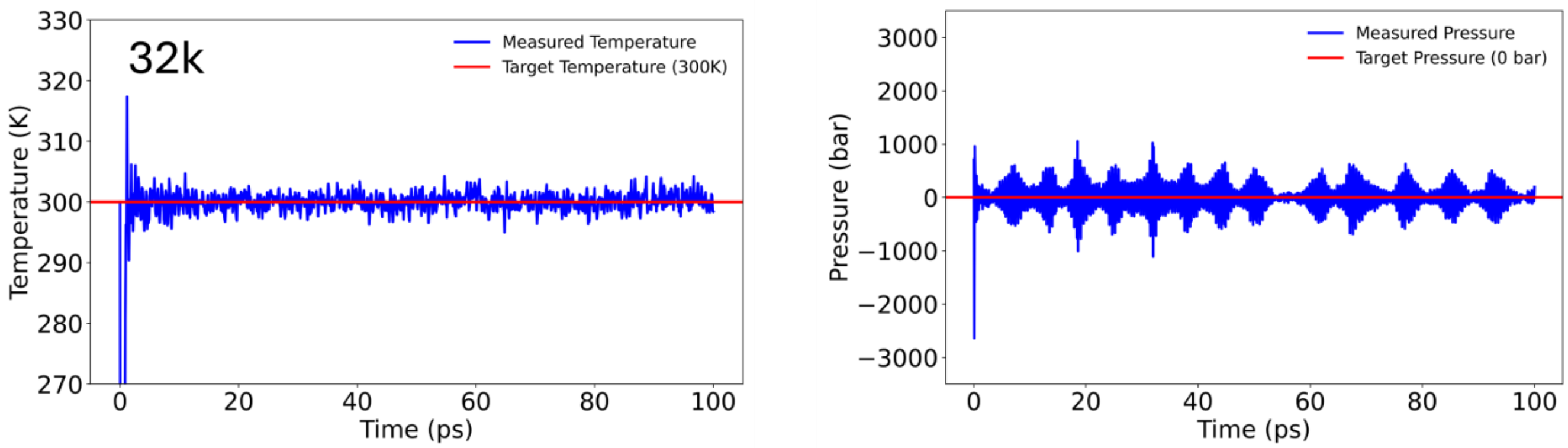

Fig. S2. Time evolution of temperature (left) and pressure (right) during the molecular-dynamics simulation of the equilibration stage for the 32k-atom Ti model. Both quantities fluctuate around the target values of 300 K and 0 bar, indicating that thermodynamic equilibration is achieved prior to compression for the simulated system.

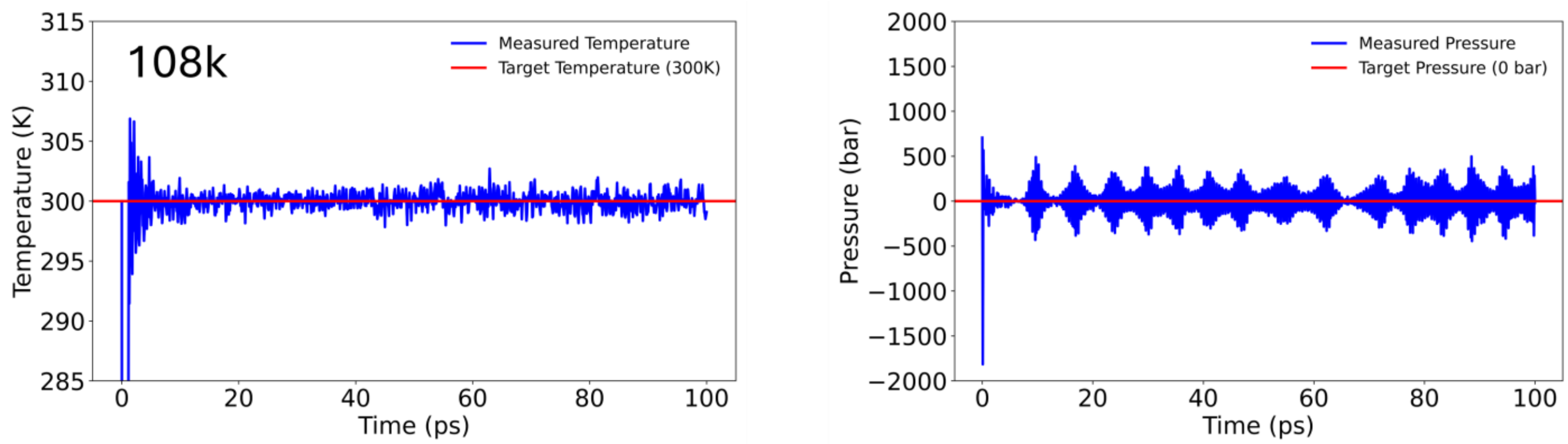

Fig. S3. Time evolution of temperature (left) and pressure (right) during the molecular-dynamics simulation of the equilibration stage for the 108k-atom Ti model. Both quantities fluctuate around the target values of 300 K and 0 bar, indicating that thermodynamic equilibration is achieved prior to compression for the simulated system.

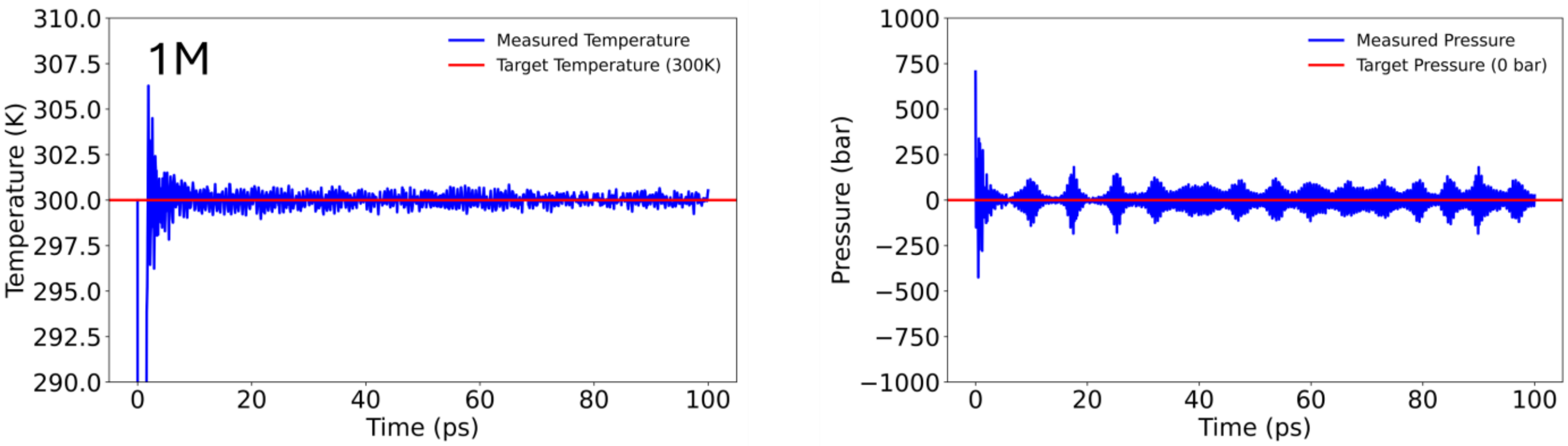

Fig. S4. Time evolution of temperature (left) and pressure (right) during the molecular-dynamics simulation of the equilibration stage for the 1M-atom Ti model. Both quantities fluctuate around the target values of 300 K and 0 bar, indicating that thermodynamic equilibration is achieved prior to compression for the simulated system.

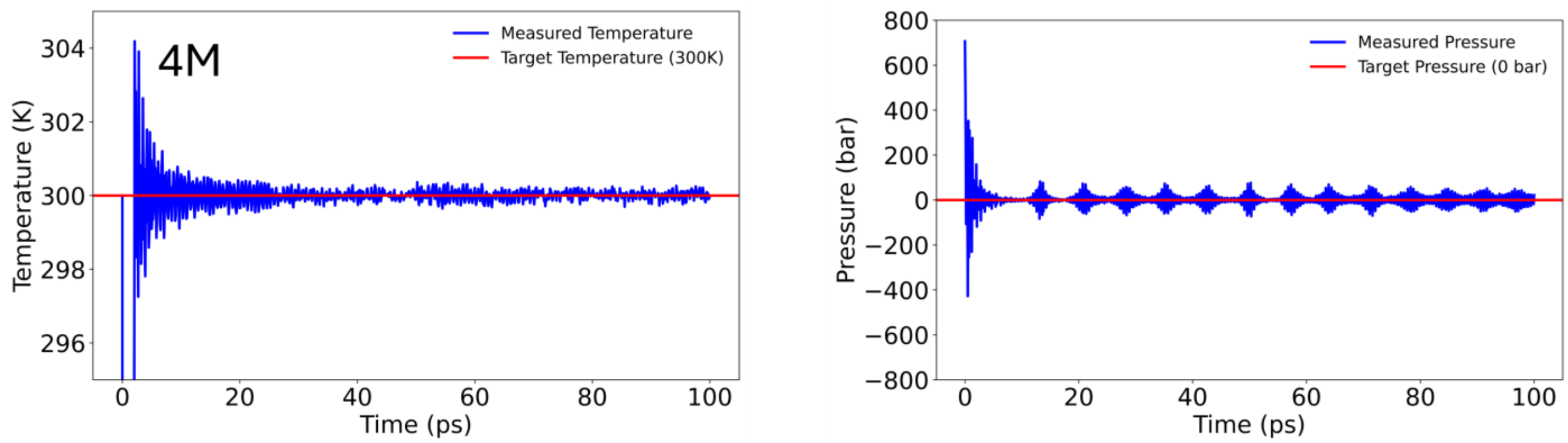

Fig. S5. Time evolution of temperature (left) and pressure (right) during the molecular-dynamics simulation of the equilibration stage for the 4M-atom Ti model. Both quantities fluctuate around the target values of 300 K and 0 bar, indicating that thermodynamic equilibration is achieved prior to compression for the simulated system.

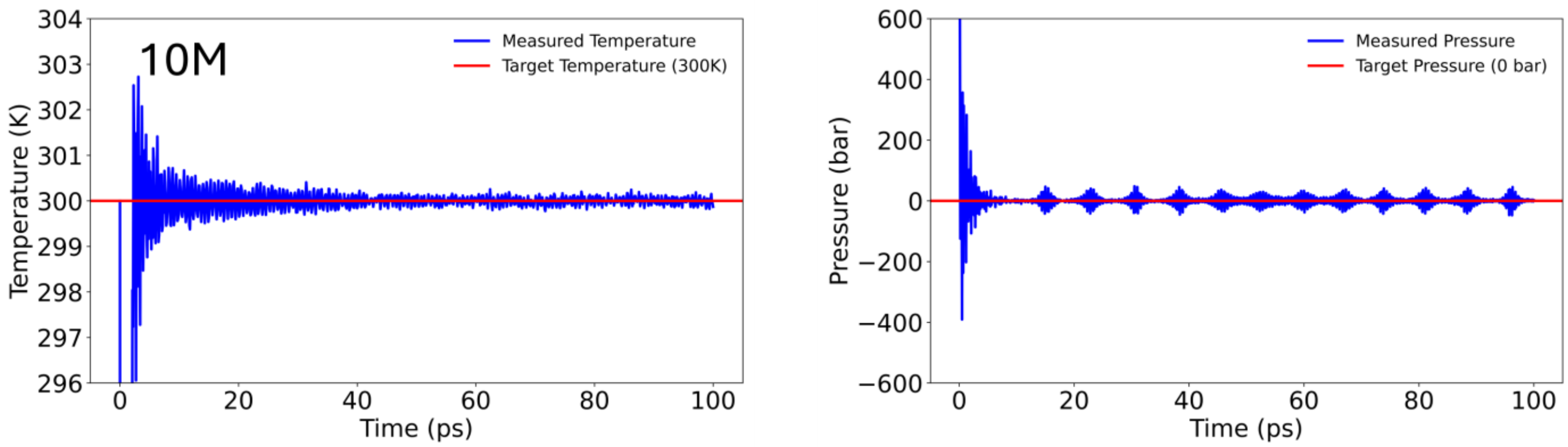

Fig. S6. Time evolution of temperature (left) and pressure (right) during the molecular-dynamics simulation of the equilibration stage for the 10M-atom Ti model. Both quantities fluctuate around the target values of 300 K and 0 bar, indicating that thermodynamic equilibration is achieved prior to compression for the simulated system.

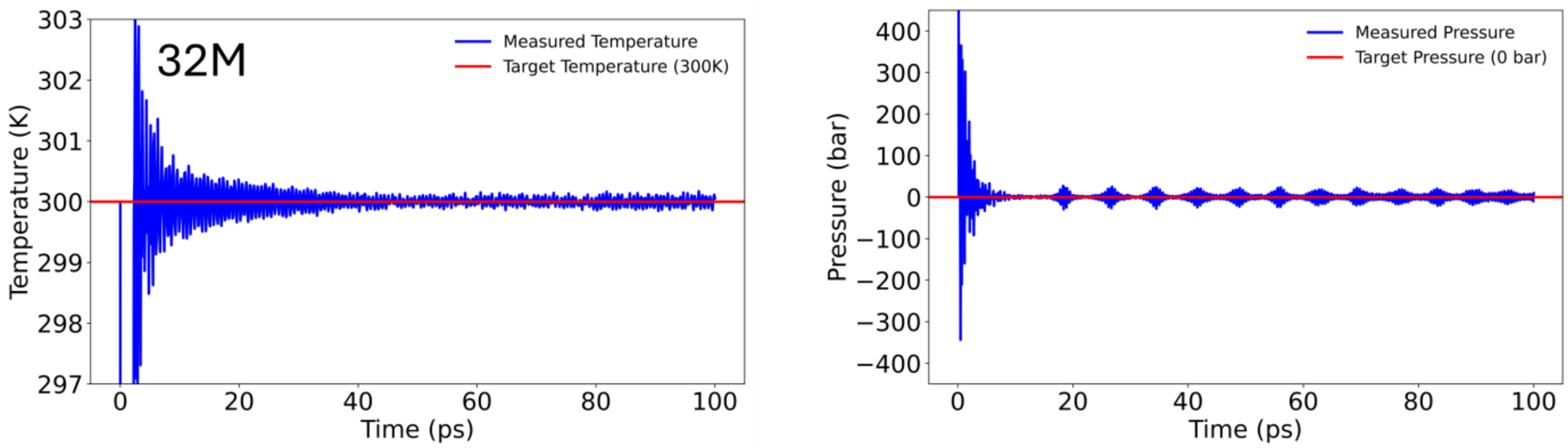

Fig. S7. Time evolution of temperature (left) and pressure (right) during the molecular-dynamics simulation of the equilibration stage for the 32M-atom Ti model. Both quantities fluctuate around the target values of 300 K and 0 bar, indicating that thermodynamic equilibration is achieved prior to compression for the simulated system.

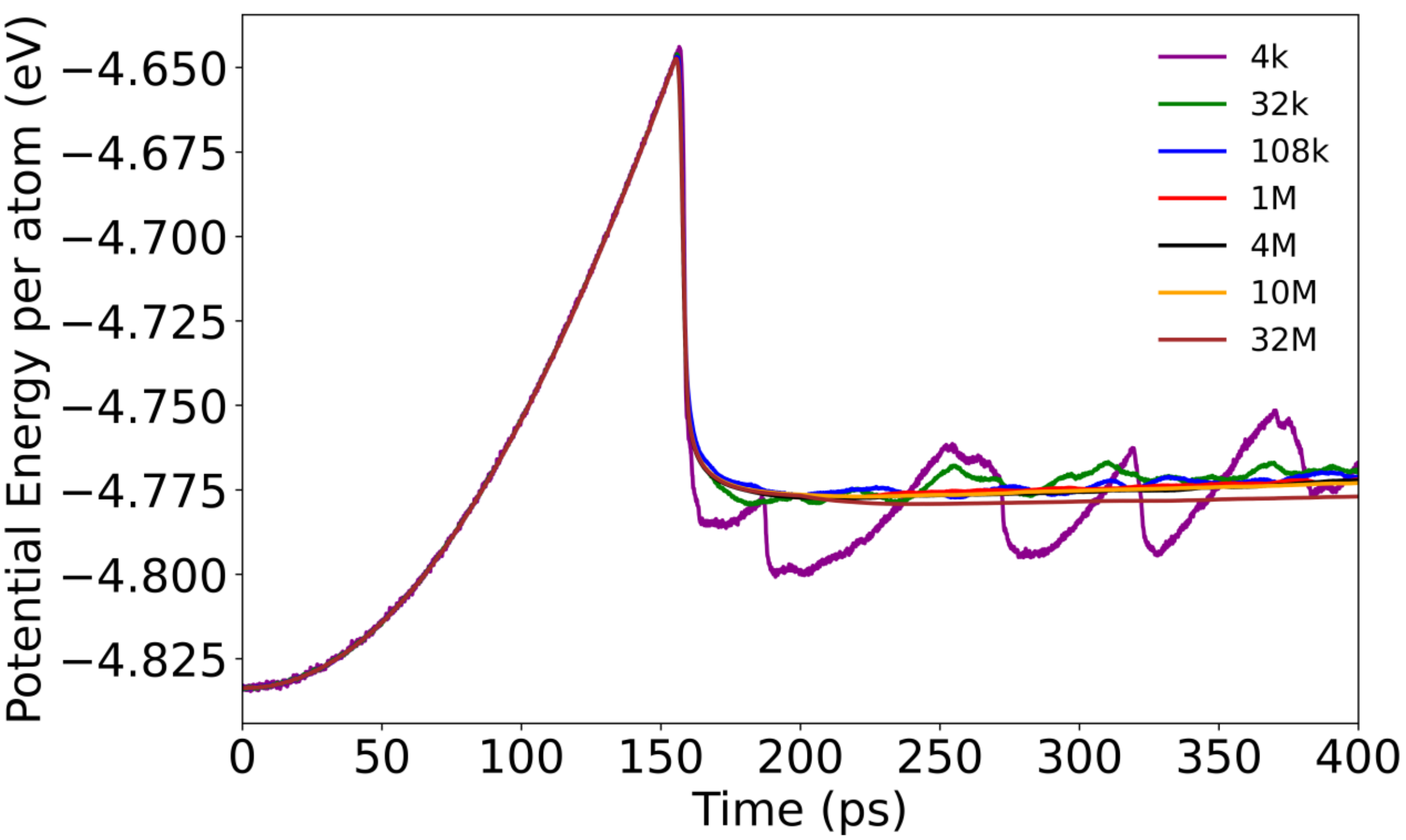

Fig. S8. Time evolution of the total potential energy (per atom) during the compression simulations at a $10^9$ $s^{-1}$ strain rate for all the different Ti modelled system sizes.

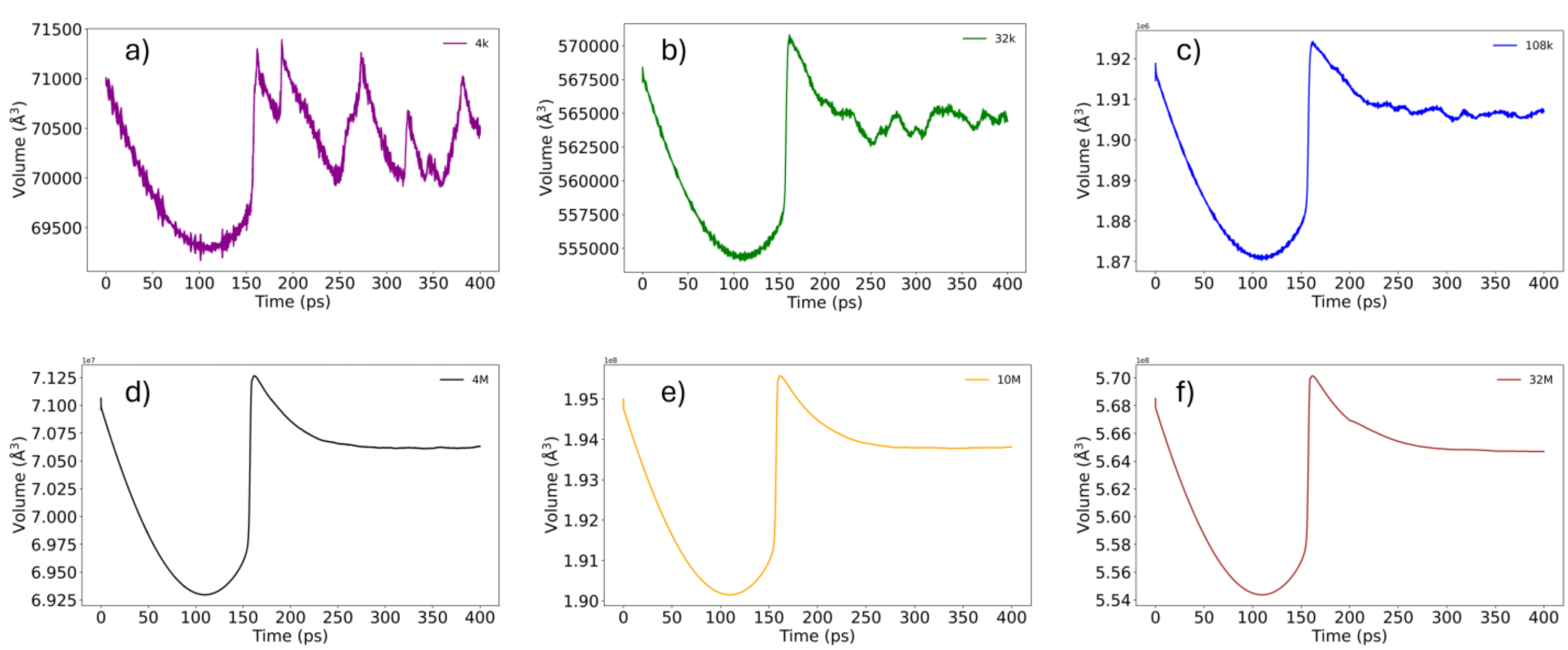

Fig. S9. Time evolution of the simulation cell volume during the compression of different Ti modelled structures at a $10^9$ $s^{-1}$ strain rate: a) 4k, b) 32k, c) 108k, d) 4M, e) 10M, and f) 32M.

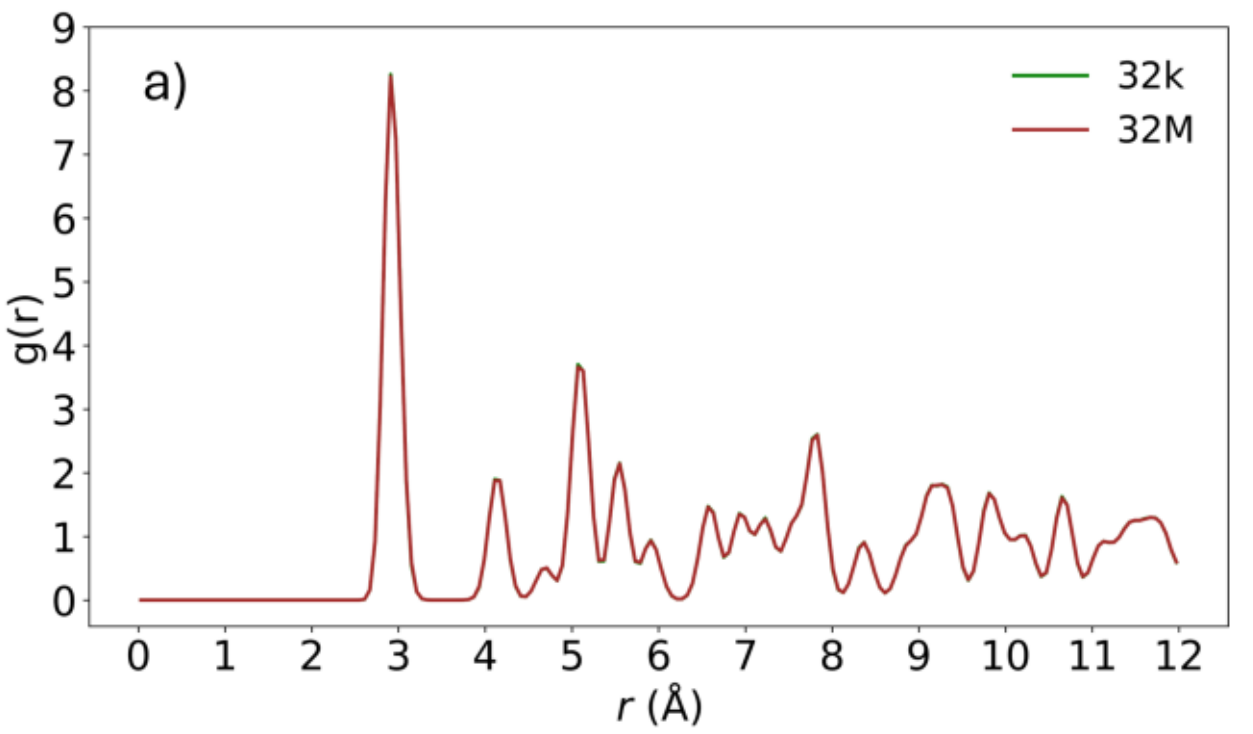

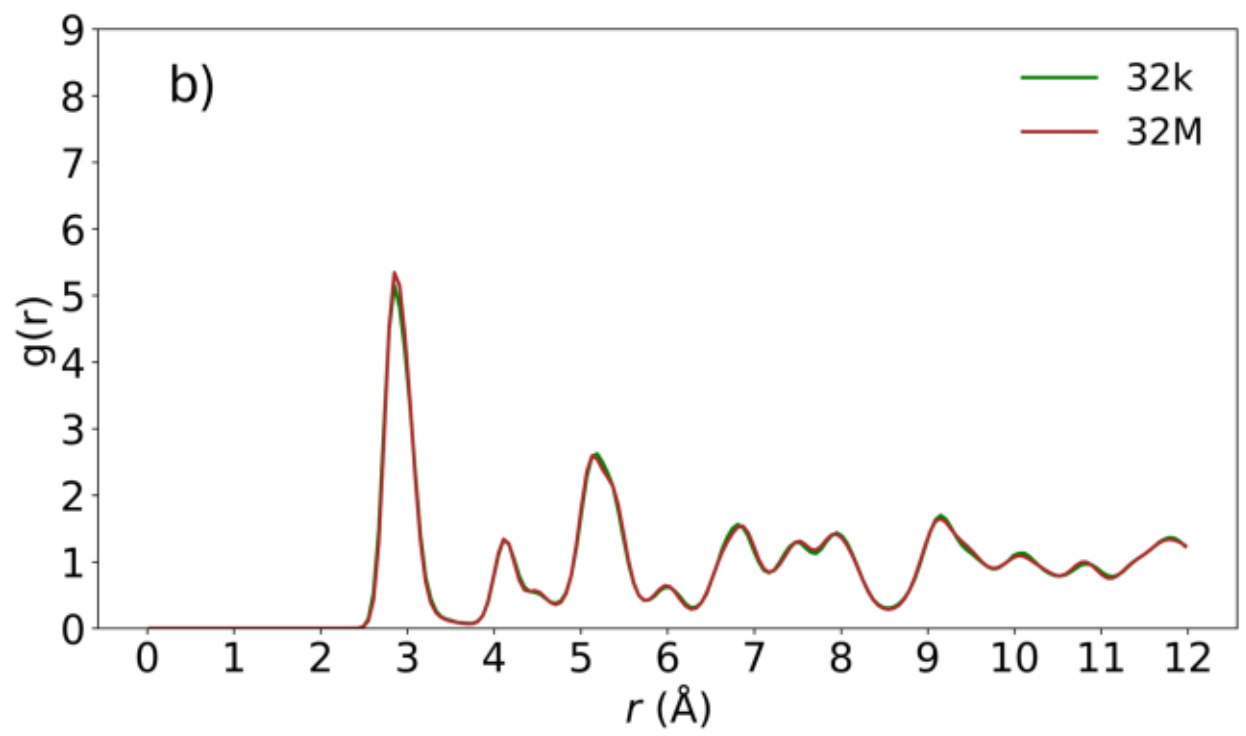

Fig. S10. Radial distribution functions, g(r), of the 32k-atom and 32M-atom Ti modelled structures: a) before compression; and b) after compression (corresponding to the simulation at the strain rate of $10^9$ $s^{-1}$, and for 40% applied strain).

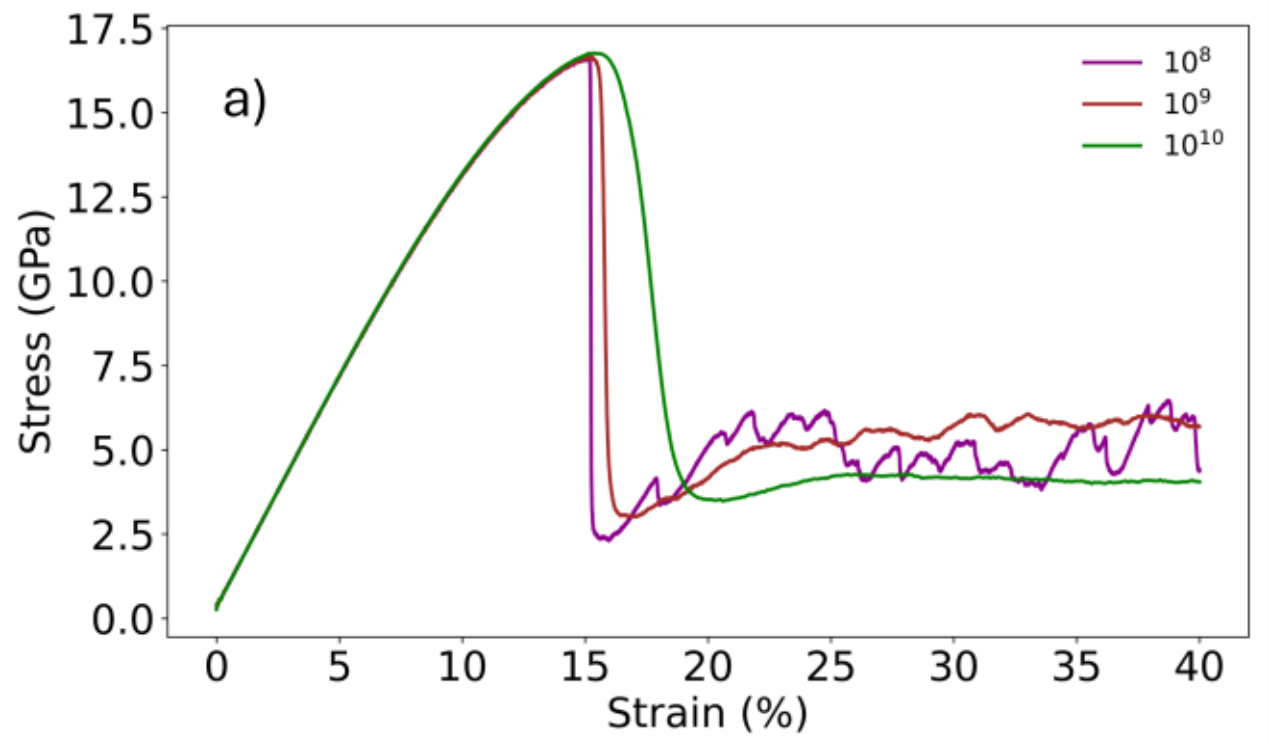

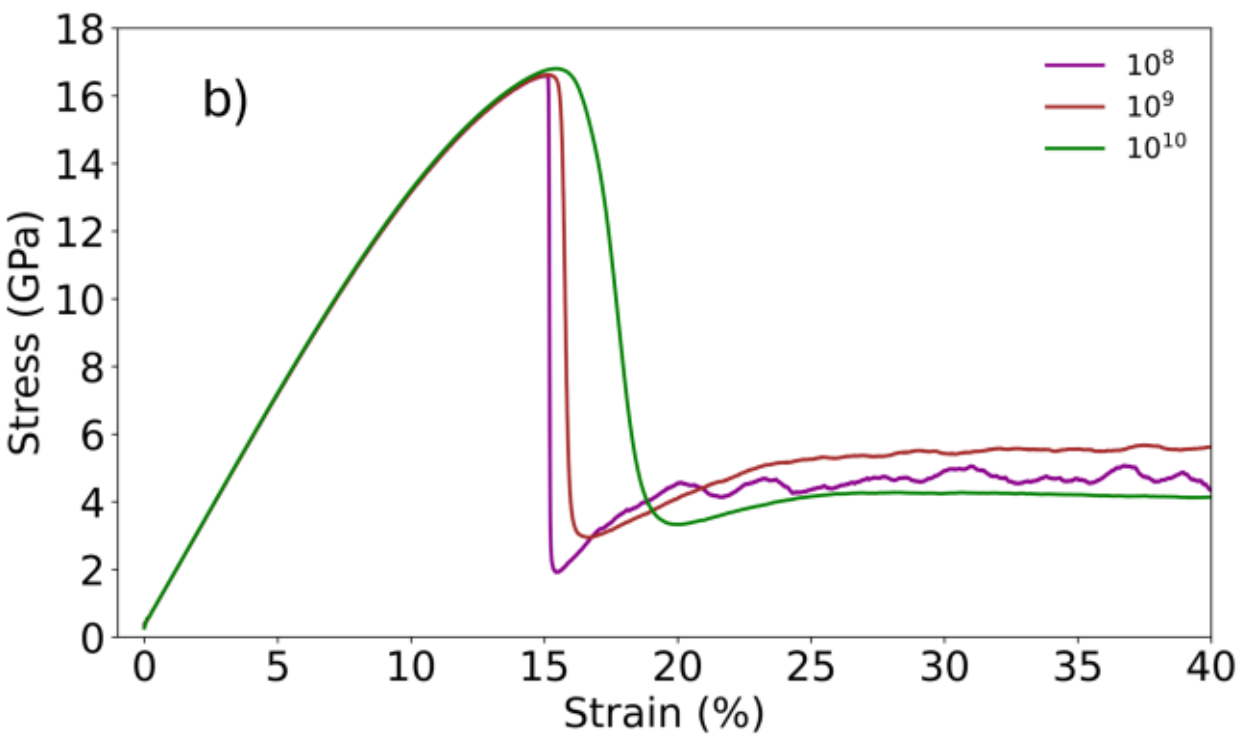

Fig. S11. Stress-strain analysis at three different strain rates ($10^8$, $10^9$, and $10^{10}$ $s^{-1}$) for two different simulated Ti system sizes: a) 108k model; and b) 1M model.